\documentclass[
11pt]{article}
\usepackage{fancyhdr}
\def\eqalign#1{\null\,\vcenter{\openup\jot
  \ialign{\strut\hfil$\displaystyle{##}$&$\displaystyle{{}##}$\hfil
      \crcr#1\crcr}}\,}

\usepackage{color}
\usepackage{hyperref}
\usepackage{graphicx}
\usepackage{dcolumn}
\usepackage{bm}
\usepackage{verbatim}

\def\iniz{\setcounter{equation}{0}{%
\rhead{\thepage}\lhead{{{{\small\bf\thesection:}
\small \ \SEC\ \  \today}}}}}

\let\a=\alpha   \let\g=\gamma  \let\d=\delta 
       
\let\m=\mu    \let\n=\nu             
\let\s=\sigma \let\t=\tau    
     
 \let\D=\Delta

\def\V#1{{\bf#1}}\def\rhs{{\it r.h.s.}\ }
\def\*{\vskip 3mm}\def\0{\noindent}
\def\be{\begin{equation}}
\def\ee{\end{equation}}
\def\bea{\begin{eqnarray}}
\def\eea{\end{eqnarray}}

\let\dpr=\partial
\def\ie{{\it i.e.}\ }
\def\eg{{\it e.g.}\ }
\def\Eq#1{\label{#1}}
\def\equ#1{(\ref{#1})}
\def\lis#1{\overline#1}
\def\defi{{\buildrel def\over=}}

\def\otto{\,{\kern-1.truept\leftarrow\kern-5.truept\to\kern-1.truept}\,}

\def\tende#1{\,\vtop{\ialign{##\crcr\rightarrowfill\crcr
 \noalign{\kern-1pt\nointerlineskip} \hskip3.pt${\scriptstyle
 #1}$\hskip3.pt\crcr}}\,}
\def\ie{{\it i.e.\ }}

\def\W#1{#1_{\kern-3pt\lower6.6truept\hbox to 1.1truemm
{$\widetilde{}$\hfill}}\kern2pt\,}

\newdimen\xshift \newdimen\xwidth \newdimen\yshift \newdimen\ywidth
\def\ins#1#2#3{\vbox to0pt{\kern-#2pt\hbox{\kern#1pt #3}\vss}\nointerlineskip}

\def\eqfig#1#2#3#4#5{
\par\xwidth=#1pt \xshift=\hsize \advance\xshift
by-\xwidth \divide\xshift by 2
\yshift=#2pt \divide\yshift by 2%
{\hglue\xshift \vbox to #2pt{\vfil
#3 \includegraphics{#4.eps}
}\hfill\raise\yshift\hbox{#5}}}
\def\Ba   {{\mbox{\boldmath$ \alpha$}}}

\def\Bn   {{\mbox{\boldmath$ \nu$}}}

\def\BF   {{\mbox{\boldmath$ \Phi$}}}

\def\BDpr {{\mbox{\boldmath$ \partial$}}}

\begin{document}

\kern-1cm
{\color{red}
\centerline{\Large\bf Equivalence of Non-Equilibrium Ensembles and Representation }\centerline{\Large \bf of Friction in Turbulent Flows: The Lorenz 96 Model}} \*

{\color{blue}%

\centerline{}
\centerline{\tt Giovanni Gallavotti}
{\color{blue}\centerline{\small{Department of Physics, University of Rome La Sapienza}}\centerline{\small Department of Mathematics, Rutgers University, Rutgers, USA}} 
\centerline{}
\centerline{\tt Valerio Lucarini}
{\color{blue}\centerline{\small Institute of Meteorology, Klimacampus, University of Hamburg,Hamburg, Germany}\centerline{\small Department of Mathematics and Statistics, University of Reading, Reading, UK}\centerline{\small Walker Institute for Climate Change Research, University of Reading, Reading, UK
}
} {\color{red}\centerline{\today} }
}

%
%

\* \0{\bf Abstract: \it} We construct different equivalent non-equilibrium
statistical ensembles in a simple yet instructive $N$-degrees of freedom
model of atmospheric turbulence, introduced by Lorenz in 1996. The vector
field can be decomposed into an energy-conserving, time-reversible part,
plus a non-time reversible part, including forcing and dissipation. We
construct a modified version of the model where viscosity varies with time,
in such a way that energy is conserved, and the resulting
dynamics is fully time-reversible. For each value of the forcing, the
statistical properties of the irreversible and reversible model are in
excellent agreement, if in the latter the energy is kept constant at a
value equal to the time-average realized with the irreversible model. In
particular, the average contraction rate of the phase space of the
time-reversible model agrees with that of the irreversible model, where
instead it is constant by construction. We also show that the phase space
contraction rate obeys the fluctuation relation, and we relate its finite
time corrections to the characteristic time scales of the system. A local
version of the fluctuation relation is explored and successfully checked.
The equivalence between the two non-equilibrium ensembles extends to
dynamical properties such as the Lyapunov exponents, which are shown to
obey to a good degree of approximation a pairing rule. These results have
relevance in motivating the importance of the chaotic hypothesis. in
explaining that we have the freedom to model non-equilibrium systems using
different but equivalent approaches, and, in particular, that using a model
of a fluid where viscosity is kept constant is just one option, and not
necessarily the only option, for describing accurately its statistical and
dynamical properties.

\* \0{Key words: \it Equivalent Equations, Turbulence, Chaotic
  Hypothesis, Fluctuation Theorem, Geophysical Flows, Parametrization,
  Lyapunov Exponents}

\* 
\def\SEC{Introduction}
\section{\SEC}\iniz\label{sec1}

Non-equilibrium statistical mechanical systems reach a steady state, after
transients have died out, with a statistical balance between forcing and
dissipation. The forcing is typically performed by an external field, while
the dissipation is taken care of by a suitably defined thermostat, which
has the role of removing the excess of energy accumulated in the system as a
result of the work due to the external field and the related processes
cascading from there. 

When considering a macroscopic description of a physical system, friction
plays the role of the thermostat. The most common way of introducing it  is
phenomelogically described by a force opposing motion proportional to a
\textit{friction constant} $\n$. The introduction of friction in this form
leads to a fundamental change in the equation of motion of the system, as
the time reversal symmetry is broken. This cannot be a fundamental model of
the process of thermostatting because the basic equations of Physics are
time reversal invariant, \ie invariant under a transformation $I$ which
anticommutes with the time evolution map $x\to S_t x$, in the sense that
$S_t I x=I S_{-t}x$ (which usually is just a velocity reversal, but 
might be more involved, \cite{Ga998}.

Hence, it is worth investigating whether the statistical, macroscopic
properties of a physical system whose microscopic dynamics obeys a simple
phenomenological friction law of the kind described above can be
equivalently described by different microscopic equations, constructed in
such a way that fundamental time reversal symmetry is preserved
\cite{Ga997b,Ga996a,Ga000a}. The main reason for studying this is not the
hope for simpler equations, but rather the possibility of having an
alternative view of the dynamics, which could reveal certain features of
the problem.

A compelling analogy can be found with equilibrium statistical mechanics,
and in particular with the concept of \textit{equivalence of the
  ensembles}, which suggests, \textit{e.g.} that the same system can be
equivalently described by the microcanonical or canonical
ensembles. Obviously, the equivalence does not extend to all properties: If
one is interested in energy (temperature) fluctuations, it would be futile,
by construction, to use the microcanonical (canonical) ensemble. A similar
discussion has not yet been systematically carried over in the context of
nonequilibrium. An early attempt at a check of the ideas in the works cited
above can be found in \cite{GRS004}.  For instance if interest focuses on
dissipation fluctuations a model in which the dissipation is a fixed
constant may affect deeply dissipation itself (obviously), and other
observables.

Since reasonable Physics will not dismiss dissipative equations, like the
Navier-Stokes (NS) equations, the approach that should be taken is to
regard the stationary {states} reached in models with equations in which
friction is constant as \textit{ensembles} whose properties can also be
described by equivalent ensembles, which are stationary states of other
equations. In particular, our goal is to construct such ensembles from
evolution equations obeying time reversal symmetry. 

In this paper a detailed investigation is pursued on a specific model and
the result suggest a very general picture, whose foundation was laid 
{out} in
\cite{Ga997b,Ga996a}. The conceptual frame is the \textit{chaotic hypothesis}
  which leads to the emergence of various properties making use of the
  general theory of chaotic motions initiated in \cite{Lo963,RT971} and
  allows deriving \textit{parameter free} predictions on various fluctuations
  like the \textit{fluctuation theorem} \cite{GC995}.  

It is important to keep in mind that the representation of dissipative
processes in \textit{e.g.} fluids, except in the rare cases where direct
numerical simulation down to the Kolmogorov scale is feasible and one can
use molecular diffusivities and viscosities, requires considerable
theoretical efforts. In fact, one must find simplified yet accurate methods
for dealing with the unresolved scales of motion and for representing
correctly complex cascades of quantities like energy and enstrophy.  The
formulation of methods for performing large eddy simulations
\cite{Sma93,Sag006,Gar009}, and, more in general, for providing a closure
to Reynolds stresses \cite{Pop000} provide prototypical examples in this
direction. The aim of our work is, however, to study in a simple example
the fundamental role of time reversibility in dissipative phenomena, and
shows, at least in the model considered, how to avoid using equations
breaking time reversal invariance without loss of information. Hence it
addresses specifically time reversal symmetry (and possibly other symmetry
properties), unlike the previous literature in the field, as the just works
cited above, which model dissipation phenomenologically with irreversible
equations. A notable exception to this approach in the existing literature
can be found in \cite{SJ993} where it is shown, for the first time, that is
possible to reproduce the properties of a dynamical system where a constant
friction plays a key role using a model obeying time reversibility.

\subsection{Equivalence of Ensembles}\label{equivalence}
Let us consider a dynamical system with $N$ degrees of freedom
\be\dot x_j= f_j(x)+ F_j- \n (L x)_j,\qquad \n>0,\ j=1,\ldots,N
\Eq{e1.3}\ee
where $L$ is a positive definite dissipation
matrix: \eg in many interesting cases 
and $(Lx)_j=x_j$ and $\n>0$, and $f(x)=-f(-x)$ (time
reversibility).

Let $E(x)$ be an observable such that $\sum_{j=1}^N \dpr_j E(x)
(Lx)_j=M(x)$ is positive for $x\ne0$. In the simple case where $L$ is
  the identity, taking $E(x)=\frac12\sum_j x_j^2\equiv x^2$, we have $M(x)=
  x^2$; then the equation
\be\dot x_j= f_j(x)+ F_j- \a(x) (L x)_j,\qquad \a(x)\defi\frac{\sum_{j=1}^N
F_j\dpr_j E}{M(x)},\Eq{e1.3A}\ee
will admit $E(x)$ as an exact constant of motion, and, if $E(x)=E(-x)$, it
will be time-reversible {with time reversal $x\to-x,\,t\to-t$.} The quantity $\alpha(x)$ will fluctuate in time and, in general, will not have a definite sign. We say
that the stationary distributions of the Eq.\equ{e1.3},\equ{e1.3A} define
\textit{corresponding ensembles} of statistical distributions.

A natural generalization to cases with several friction constants is
obtained if $L_\Bn$ is a positive matrix depending linearly on $k$ positive
friction constants $\Bn=(\n_1,\ldots,\n_k)$, $(L_\Bn x)_j=
\sum_{j'=1}^N\sum_{s=1}^k \n_s L^s_{j,j'}x_{j'}$: then the stationary
distributions for the equations

\be\eqalign{&\dot x_j=f_j(x)+F_j- (L_\Bn x)_j,\qquad {\rm and}\quad \dot
  x_j=f_j(x)+F_j- (L_\Ba x)_j, \cr}\Eq{e1.3B}\ee 
will be said to form two \textit{corresponding ensembles} of statistical
distributions with dissipation balanced on $k$ observables $\V
E(x)=(E_1(x),\ldots,E_k(x))$, with $E_j(x)=E_j(-x)$, if the $k\times k$
matrices

\be M_{r,s}(x)\defi \sum_{j,j'=1}^N \dpr_j E_r(x)L^s_{j,j'}x_{j'}\Eq{e1.3C}
\ee
are positive definite for $x\ne0$, and if the coefficients
$\Ba(x)=(\a_1(x),\ldots,\a_k(x))$ are defined as

\be \eqalign{
&\a_r(x)=\sum_{s=1}^k
M(x)^{-1}_{rs} \sum_{j=1}^N \dpr_j E_s(x) f_j(x),\cr
&\Ba=M(x)^{-1}\BF(x),\qquad \BF_r(x)=(\V F(x)\cdot{\BDpr
E_r(x) })\cr}\Eq{e1.3D}\ee
The key property of the equations \equ{e1.3D} is that $\V E(x)$ are $k$
exact constants of motion; furthermore the equation is time reversal
invariant (with time reversal symmetry given by {$I: x\rightarrow -x,
t\rightarrow -t$}), as opposed to Eq.(\ref{e1.3}). 

The equations Eq.\equ{e1.3B} will be said to have `\textit{dissipation balanced on 
the observables $\V E$'}.

A simple class of equations with dissipation balanced on $\V E(x)$ is
obtained in the cases in which $\V E(x)$ are quadratic positive definite
constants of motion for the inviscid, not forced equation $\dot x=f(x)$ and
the viscous equations are

\be \dot x_j=f_j(x)+F_j(x)-\sum_{r=1}^k\n_r\dpr_j E_r(x)\Eq{e1.3E}
\ee
In this case the $\BF(x)$ does not depend explicitly on $f$.
However the balanced dissipation equations are more general.

In \cite{Ga997b,Ga002}, it was proposed, in special cases and {for
  strong forcing}, that given a dissipative equation, the constant friction
vector $\Bn$ could be replaced by the corresponding time reversible
equations with a variable vector $\Ba=(\a_1,\ldots,\a_k)$ as defined above
which obeys time reversal symmetry, as opposed to the original
dissipative-balanced system.

As a notable example, the incompressible {$2$-dimensional Navier-Stokes
  (NS) equation in a periodic container} can interpreted, {according to
  Eq.\equ{e1.3}, as describing a dissipative-balanced system:
\be\dot{\V u}+(\V u\cdot \BDpr)\V u=-\BDpr p+ \V g+\n\D\V u =0,\quad \BDpr
\cdot\V u=0\Eq{e1.4}\ee
because $E(\V u)=\frac12 \int (\BDpr \V u)^2(x)dx$, being the dissipation
of kinetic energy (apart from a factor $1/\nu$), is a constant of motion in
$2D$ incompressible fluids with periodic boundary conditions and $\frac{\d E(\V
  u)}{\d u_j(x)}=\int \sum_{rk}\dpr_r u_k \d_{jk}\dpr_r\d(y-x)\,dx\equiv-\D
u_j(x)$. The equivalent model is then:}
\be\eqalign{
&\dot{\V u}+(\V u\cdot \BDpr)\V u=-\BDpr p+ \V g+\a(\V u)\D\V u ,\qquad
\BDpr \cdot\V u=0
\cr
&\a(\V u)\defi \frac{
-\int \V g\cdot\D \V u\,d\V x+
\int \D\V u\cdot\Big((\W u\cdot\W\dpr)\V u\Big)\,d\V x}{\int (\D \V u)^2}
\cr
}\Eq{e1.7}\ee
which via the spatial Fourier decomposition of the velocity field $\V u$
becomes:
\be\eqalign{
\a(\V u)\defi&
\frac{\sum_{\V k} \V k^2\,\V g_{\V k}\cdot \V u_{-\V k}}{\sum_{\V k} \V k^4|\V
    u_{\V k}|^2}, \qquad D=2\cr
}
\Eq{e1.7b}\ee
In dimension $3$ the same fluid with dissipation balanced on the
vorticity observable $E(\V u)$ cannot be regarded as an element of the
  special class given in Eq. \ref{e1.3E} where the observables $\V E$ are constants of motion \textit{and} the viscous term contains their derivatives, because
in this case $E(\V u)$ in absence of friction $E(\V u)$ is not
conserved. If one considers instead as conserved quantity for the unviscid  flow $K(\V u)=\int \V u(x)^2dx$, the corresponding irreversible equations (both in 2 and 3 dimensions) of the form given in Eq. \ref{e1.3E} are the the so-called Ekman friction equations, where the dissipative term $\D\V u$ of the Navier-Stokes equations is substituted by a term of the form $- \V u$. 

The $3D$ fluid can nevertheless be considered to have dissipation balanced
on the vorticity as in Eq.\equ{e1.7}: in this case the inertial term will
appear explicitly in the definition of $\a(x)$, so that $\a(x)$ will be
given by Eq.\equ{e1.7} and it will not be as simple as Eq.\equ{e1.7b}.
It will be
\be
\a(\V u)\defi\frac{\sum_{\V k} \V k^2\,\V g_{\V k}\cdot \V u_{-\V k}
+\V k^2 \V u_{-\V k}\cdot\sum_{\V h} (\W u_{\V k-\V h}\cdot i\W h)\V u_{\V h}}
{\sum_{\V k} \V k^4|\V
    u_{\V k}|^2}, \qquad D=3\Eq{e1.7c}\ee
The equations Eq.(\ref{e1.7}),(\ref{e1.7b}) have the
 property of having $E_j(x)$ $\forall j =1 ,\ldots,k$, or, respectively,
 $E(\V u)$, exactly constant.

The proposed equivalence between the balanced dissipative equations and the
corresponding reversible equations is articulated in several properties of
their stationary distributions, reminiscent of the equivalence statements
between different ensembles in statistical mechanics.

Let $\m^i_{\Bn}$ be the stationary distribution for the irreversible
balanced dissipative system with friction coefficients $\Bn$ and
$\m^r_{\lis{\V E}}$ the stationary distribution for the reversible equation
with $\V E$ as {a constant} of motion vector with value $\lis{\V E}$.
{\it Then $\m^i_\Bn$ will be said to correspond to $\m^r_{\lis{\V E}}$
  if $\m^i_\Bn({\V E})=\lis\V E$ or (equivalently, as we shall argue) if
  $\m^r_{\lis {\V E}}(\Ba)=\Bn$}.

The correspondence that we discuss will be interpreted to imply
equivalence: \ie that setting the value of the viscosity coefficient $\Bn$
is conceptually equivalent to setting the value of the physical quantities
${\V E}$, {at least in the limit of large forcing}. Therefore, it will
become possible, at least in the cases and under the assumptions discussed
in the following, to take two different, but equally effective, points of
view for studying the properties of the considered system {in the same
  sense in which different equilibrium ensembles become equivalent in the
  \textit{thermodynamic limit} (which in most cases is reached already in not
  too large systems, as shown by the simulations with few degrees of
  freedom)}.

The two viewpoints differ microscopically in term of {the} representation
of reversibility, but provide the same statistical mechanical picture.

More precisely some among the equivalence properties that we consider are:\*

\0{\bf(1)} If $\m_{\lis{\V E}}(\Ba)=\Bn$ then $\m_{\Bn}(\V E)=\lis{\V E}$:
the equivalence is reflexive.  
\\
{\bf(2)} If $g(x)$ is a smooth observable
in a large class (for instance in the case of the NS equations if its
Fourier transform can be expressed in terms of few low Fourier modes), the
statistics of $\g\defi \frac1T\int_0^T g(S_t x)dt$ is the same in the two
systems for $T$ large \\
{\bf(3)} The fluctuation relation, { \cite{GC995}}, holds for
the dissipation function
\be
\s(x)\defi\sum_j \dpr_j[(\dpr_j\V E(x))\cdot
    M^{-1}(x)\BF(x)]\Eq{e1.8}
\ee
which in the example of Eq.\equ{e1.7} {and for $\V g_{\V k}\ne0$ only for
few small $\V k$ (\ie for large scale forcing)} is
\be \eqalign{
\s(\V u)=&2(\sum_{\V k}\V k^2)\a(\V u)+
\frac{\sum_{\V k}\V k^4\V g_{\V k}\cdot\V u_{-\V k}}
{\sum_{\V k}\V k^4|\V u_{\V k}|^2}\cr
&
-2\frac{(\sum_{\V k} \V k^2\V g_{\V k}\cdot\V u_{-\V k})
(\sum_{\V k} \V k^6|\V u_{\V k}|^2)}
{(\sum_\V k\V k^4|\V u_{\V k}|^2)^2}\,
\simeq\, 2(\sum_{\V k}\V k^2)\a(\V u)
\cr}\Eq{e1.9}
\ee
where the sum over $\V k$ is over the modes below the ultraviolet cut-off
used.\footnote{\small In $3$ dimensions the natural cut-off would be the
  Kolmogorov scale; in $2$ dimensions the cascade is inverse and the
  interpretation is more subtle, see \cite{Ga997b}.}

This means that if $\lis\s$ denotes the average value of the dissipation
function in the stationary states $\m^i_{\lis{\V E}}$ and if
$p=\frac1\t\int_0^\t \frac{\s(S_t x)}{\lis\s}$ then the probability
$P_\t(p)dp$ of the event $p\in dp$ in the distribution $\m_{\lis{\V E}}$ is
multifractal and
\be \frac1\t\log\frac{P_\t(p)}{P_\t(-p)}=\lis\s\, p+ O(\t^{-1})\Eq{e1.10}\ee
A more mathematical statement is that the rate function $\g(p)$
describing the multifractal distribution%
\footnote{\small This is a function $\g(p)$ such that the probability the
  $p\in\D$ is {asymtpotic as $\t\to\infty$ to}
  $const\,e^{\t\max_{p\in\D}\g(p) }$: in Anosov systems it is analytic,
  \cite{Si977,GBG004}, in $p$ for $|p|<p^*$ for some $p^*$; in time
  reversal symmetric Anosov systems $p^*\ge1$.}
  of the random variable $p$ with respect to $\m_{\lis{\V E}}$ has the
  symmetry property $\g(-p)=\g(p)-\lis\s \,p$ for $|p|<p^*$ where $p^*$ is
  a quantity depending on the system, however $p^*\ge1$, for details see
  \cite{Ga995b}.  \\ 
\0{\bf(4)} The Lyapunov spectra of the two
  distributions coincide.  \*

\subsection{Our Model}

The goal of this paper is to substantiate the conjecture of equivalence
between an irreversible dissipative-balanced model and the corresponding
reversible model, constructed as detailed above. We consider the so-called
Lorenz '96 model, proposed by E. Lorenz in a series of papers
\cite{LE998,Lo005}:
\be\dot x_j=x_{j-1}(x_{j+1}-x_{j-2})+F-\n x_j, \qquad j=0,\ldots,N-1
\Eq{e1.1}\ee
and periodic conditions $x_k=x_{N+k}$, $\forall k$. The \rhs of the
equations has the typical structure of a sum of an energy conserving and
time reversal invariant part plus a forcing part plus a dissipation
violating time reversal (which in this case is the map $(I x)_j\defi
-x_j$). 

The variables $x_j$ of the model \ref{e1.1} have been originally
interpreted as the value of a generic meteorological quantity measured
around a latitudinal circle at a regular longitudinal interval. 

The system shows chaotic motions, with a substantial fraction of positive
Lyapunov exponents, already at moderate forcing \ie at values of
$R=\frac{F}{\n^2} \ge8$: via the scaling $x_j(t)=\n X_j(\n t)$ the equation
acquires the dimensionless form
\be \dot X_j=X_{j-1}(X_{j+1}-X_{j-2})+R-X_j, \qquad j=0,\ldots,N-1
\Eq{e1.2}\ee
The model is has become standard test-bed for predictability studies
\cite{Orr03}, for devising advanced data assimilation techniques
\cite{Tre04,Tre10} and for constructing new parametrizations \cite{Wil05},
so that today it has become \textit{the} toy-model for studies of the
nonlinear properties of geophysical flows.

Moreover, such a model has been extensively studied for constructing
response operators to perturbations \cite{majda07} and for testing
accurately Ruelle's response theory \cite{lucarini2011,Luc2014}.{The
applicability, up to an extremely high degree of precision of the Ruelle
theory, described in \cite{lucarini2011} for $R=8$ supports the idea that
the system features Axiom A-like properties \ie - is \textit{sufficiently
  chaotic} - already at moderate forcings.}

Assuming the chaotic hypothesis, \cite{GC995b,Ga013}, the system will reach
a stationary state $\m^i_R$ and $E=\frac12\sum_{j=1}^{N} X_j^2$ will be a
fluctuating variable with average value $\lis E^i_R=\int E(x)d\m(x)^i_R$.  We
will consider as relevant macroscopic observable the \textit{momentum} of
the system, defined as $M=\sum_{j=1}^{N} X_j$, which will fluctuate
around its average value $\lis M^i_R=\int M(x)d\m(x)^i_R$.

The rescaled model Eq.(\equ{e1.2}) fits into the general {dissipation
  balanced} form given in
Eq.(\ref{e1.3}) when taking $k=1$, $E(X) =\frac12\sum_j X_j^2$, $F_j=R$,
$\n=1$.  

We will test the four equivalence properties described in the previous
section on the irreversible and reversible versions of the Lorenz '96 model
for a large range of value of $R$ in the turbulent regime. We want to
emphasize that all the computations presented in this paper have ben
performed on a regular laptop computer equipped with
MATLAB\textregistered $ $ V. 7.9 using a net total of about 10 days of CPU
time.

The paper is organized as follows. In Sec.\ref{sec2} we describe
some basic properties of the irreversible Lorenz '96 model, in particular
confirming simple scaling laws describing the statistical properties of
$M$, $E$, and showing new scaling laws for the spectrum of Lyapunov exponents. In Section \ref{sec3}
we describe the properties of the reversible Lorenz '96 model, showing that
the conjectured equivalence holds to a very high precision, and in
particular we discuss the validity of the fluctuation relation. In
Sec. \ref{sec4} we draw our conclusions. Finally, in the two Appendices we provide the scripts used to perform the computations presented in this paper.

\def\SEC{Properties of the Irreversible Lorenz '96 Model}
\section{Properties of the Irreversible Lorenz '96 Model}\label{sec2}\setcounter{equation}{0}
A good starting point in the investigation of the model given in
Eq.(\ref{e1.2}) is to consider the time evolution of $E$:
\begin{equation}
\dot{E}=-2E+RM
\label{econs}
\end{equation}
which implies that at steady state 
{$\lis E^i_R=\frac12 R \lis M^i_R$}. 

The properties of Eq.(\ref{e1.2}) have been studied in detail in a number
of papers. We consider the case where $N$ is sufficiently large
(e.g. $N\geq 20$), so that it is possible to define stable (with respect to
$N$) intensive properties (see \textit{e.g.} \cite{Kar10}). Unless
otherwise stated, all presented results refer to the case $N=32$. It is
found that for $R<8/9$ the system features as unique attractor the fixed
point $X_j=R$, $\forall j$, while for larger values of $R$ {($<\sim5$)}
periodic motions due to nonlinearly equilibrated waves are observed. For $R
\geq 5$, the system is in a turbulent regime, featuring multiple positive
Lyapunov exponents. We will focus on the turbulent regime, and consider in
the rest of our analysis $R\geq 8$. Our simulations have been run for
$R=2^p$, $p=3,\ldots,{11}$. The statistics have been collected on $10^4$ time
units after discarding an initial transient. The equations have been
integrated using the MATLAB\textregistered $ $ V. 7.9 routine \texttt{ode45},
which allows for imposing a given relative and absolute precision in the
integration through adaptive {steps}. A template for the
MATLAB\textregistered$ $  scripts used for investigating the statistical
properties of $E$ and $M$ for the Lorenz '96 system is given in Appendix
\ref{scr1}. With small modifications, the scripts can be used also in Gnu-Octave (3.6.4-3). We have selected a relative and absolute precision of
$10^{-8}$, which ensures extremely high precision in the simulation. As
already observed by Lorenz \cite{Lo005}, in such a regime the long term
average of the energy $\lis E^i_R$ obeys accurately a scaling law as
follows:
\begin{equation}
\frac{\lis E^i_R}{N}\sim c_E R^{4/3} \quad \quad c_E=0.59\pm 0.01.
\end{equation}
Considering Eq. (\ref{econs}), we derive:
\begin{equation}
\frac{\lis M^i_R}{N}\sim 2 c_E  R^{1/3}.
\end{equation}
We have detected numerically several other scaling laws of considerable
interest when {studying} higher order moments of the distribution of $E$
and $M$. The first significant result refers to the standard deviation of
the energy $std(E)^i_R=\left(\overline{E^2}^i_R - (\lis
E^i_R)^2\right)^{1/2}$:
\begin{equation}
\frac{std(E)^i_R}{N}= \tilde{c}_{E}R^{4/3}, \quad \tilde{c}_{E}\sim 0.2 c_E.\label{stdE}
\end{equation}
The distribution of $E$ is to a good approximation Gaussian, and for each value of $F$ its standard deviation is proportional to the mean value.  The standard deviation of $M$ can be described for $R\geq 200$ as 
\begin{equation}
\frac{std(M)^i_R}{N}= \tilde{c}_{M}R^{2/3}, \quad \tilde{c}_{M}\sim 0.046\pm 0.001.
\end{equation}
so that in this regime $std(M)^i_R/\lis M^i_R\sim 0.039 \pm0.001$ $R^{1/3}$. Therefore, as $R$ grows, the relative size of the fluctuations of $M$ increases, so that it is more and more likely to observe negative fluctuations of $M$. As soon as $R\geq 2000$, the standard deviation of $M$ is larger than half of its mean value, and larger than the mean value itself when $R\geq 20000$. This will become relevant when studying in the next section the fluctuation relation for the reversible model.
As a last diagnostic property of the system, we note that the decorrelation
time of $M$, \textit{i.e.} the 
{time needed for the value of its autocorrelation function to reduce by a factor $e$}, obeys the following scaling relation:
\begin{equation}
t_{dec,M}\sim c_{t,M} R^{-2/3} \quad c_{t,M} =1.28 \pm0.01\label{tdecm}
\end{equation}
Quite naturally, the stronger the forcing, the more intense the turbulence, the faster the loss of memory of the system. 
\begin{figure}[ht]
\includegraphics[width=1.\columnwidth]{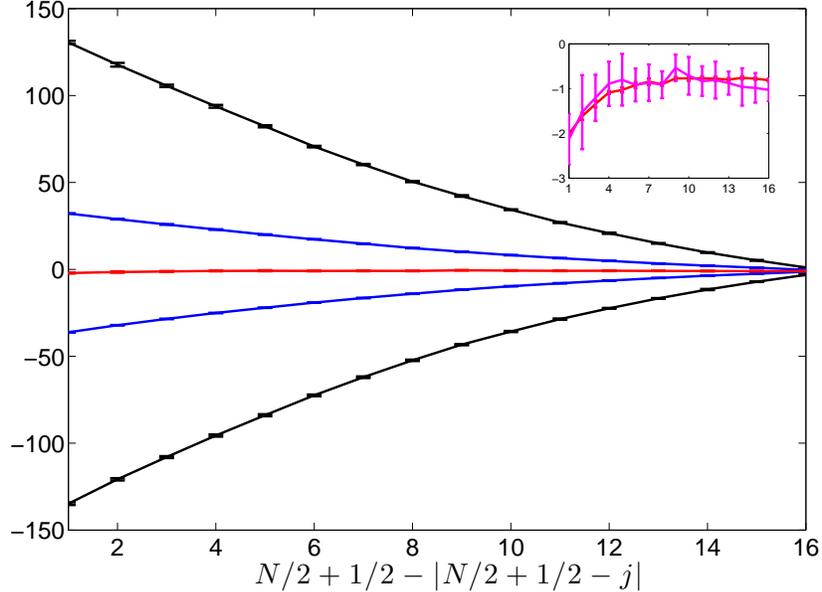}
\caption{Black line: Lyapunov exponents for $R=2048$ arranged
  pairwise. Magenta line: corresponding value of
  $\pi(j)=(\lambda_j+\lambda_{N-j+1})/2$. Blue line: Lyapunov exponents for
  $R=256$ arranged pairwise. Red line: corresponding value of
  $\pi(j)=(\lambda_j+\lambda_{N-j+1})/2$. Irreversible model.}
\label{fig1}
\end{figure}
In order to provide a synthetic picture of the dynamical properties of the
system, we next analyze its spectrum of Lyapunov exponents
\cite{eckmann85}, obtained using the classical Benettin et
al. \cite{benettin1980lyapunov} algorithm. A template for the
MATLAB\textregistered  { } scripts used for studying the Lyapunov exponents of
the Lorenz '96 system is given in Appendix \ref{scr2}. Also in this case, with small modifications, the scripts can be used also in Gnu-Octave (3.6.4-3). In Fig. \ref{fig1},
we present the spectrum of Lyapunov exponents for $R=256$ and $R=2048$ {and plot also the value of $\pi(j)=(\lambda_j+\lambda_{N+1-j})/2$}. 
  {The confidence intervals shown in Fig. \ref{fig1} indicate the $3\sigma$-uncertainty range around the expectation value. For each $\lambda_j$  and $\pi(j)$,  $\sigma$ is  computed as standard deviation of  $n=10$ estimates of  obtained running the model at steady state for $1000$ $t_{dec,M}$ time units starting from different initial conditions,  divided by the square root of $n$.  The expectation value is computed, instead, as average of the $n$ estimates.} 
 As expected, the Lyapunov exponents increase with $R$, and, remarkably, each couple of
Lyapunov exponents sums up to a value smaller in absolute value than each
of them, and smaller than zero. Let's investigate this further.

It is known that the Lyapunov exponents {$\lambda_j$} of a system
consisting of a Hamiltonian evolution plus a constant forcing plus a linear
dissipation with constant coefficient {$\n$ have the property
  $\lambda_j=-\lambda_{N+1-j}$)}, \cite{Dr988,DM996}. The time translations
of the inviscid and unforced dynamics in the \textit{r.h.s.}
Eq. (\ref{e1.2}) {are reversible (hence have at least a common property
  with the usual Hamiltonian evolutions)}, and the friction coefficient and
  the forcing are independent of $X$, it is tempting to interpret
  Fig. \ref{fig1} using such framework.

Some remarks need to be made. In \cite{Ble2013} it is explained that the
inviscid and unforced dynamics is not Hamiltonian. As a result, the
Jacobian matrices along the trajectories (which live on the spherical shell
with energy set by the initial conditions) are not simplectic. Nonetheless,
for such dynamics, we discover that the infinite time (but not the local)
Lyapunov exponents obey a Hamiltonian-like pairing rule for any given value
of the initial energy, so that $\lambda_j=-\lambda_{N+1-j}$ (not shown).



\begin{figure}[h]
\includegraphics[width=1.\columnwidth]{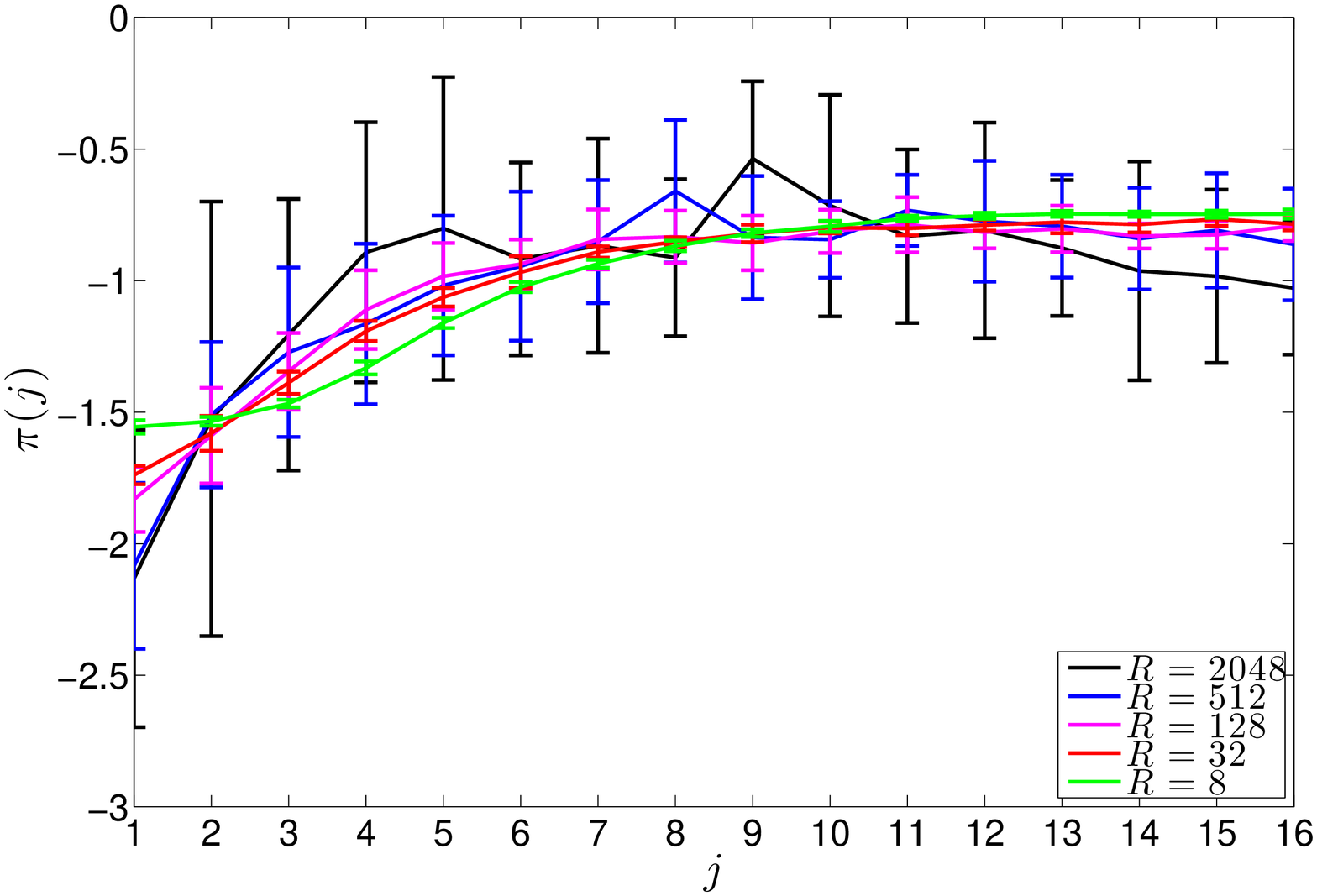}
\caption{Pairing rule $\pi(j)=(\lambda_j+\lambda_{N-j+1})/2$ for selected values of $R$.  Irreversible model.}
\label{fig2}
\end{figure}

Results are presented in Fig. \ref{fig2}, where we plot {$\pi(j)$}  {for  $R=2^{2q+1}$, $q=1,\ldots,5$, excluding the other considered values of $R$ for reasons of readability. } We
observe an emerging rule {of pairing to a quite well defined
  $R$-independent curve over a large range of values of $R$, $8\le R\le
  2048$.}\footnote{Also featuring \textit{modest} deviations , compared to
    the size of the strongly $R$-dependent Lyapunov exponents, from what
    we would have obtained (a constant -1 value) had the inviscid, unforced
    dynamics been Hamiltonian.}  {The figure shows that for all $j$'s the width of the confidence interval increases with $R$, basically because  $\pi(j)$ is constructed as algebraic sum of two quantities of opposite sign, whose average values and fluctuations increase in  magnitude with $R$.}

\begin{figure}[ht]
\includegraphics[width=1.\columnwidth]{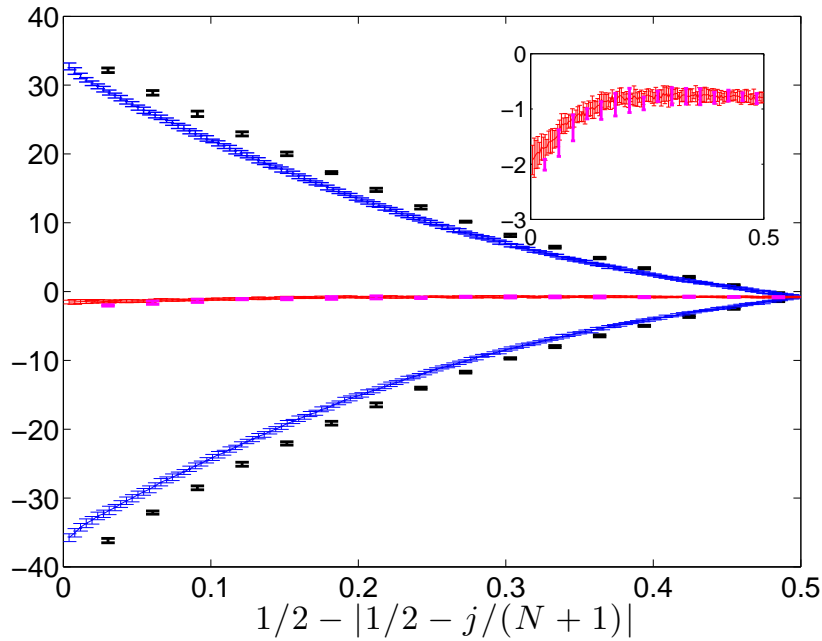}
\caption{$R=256$: Lyapunov exponents $\lambda_j$ for $N=256$ (blue line) and $N=32$ (black markers).  Values of $\pi(j)=(\lambda_j+\lambda_{N-j+1})/2$ for $N=256$ (red line) and $N=32$ (red marker); {a zoom is provided in the insert}. Irreversible model. }
\label{fig3}
\end{figure}

While the fact that observables such as ${\lis E_R/N}$ and ${\lis M_R}/N$
weakly depend on $N$ is quite intuitive, we want to substantiate the fact
that the spectrum of the Lyapunov exponents of the system does not depend
asymptotically on the value of $N$, so that   {as} $\lim N\rightarrow
\infty$ one expects, see \textit{e.g.} \cite{LPR986}, to find a density of Lyapunov
exponents. We report in Fig. \ref{fig3} the spectrum of Lyapunov exponents
obtained for $R=256$ and $N=256$, which is very similar  to
what reported in Fig. \ref{fig1} for the corresponding value of $R$ once we
rescale $j$ to $j/N+1$.  {While the confidence intervals do not overlap, we are confident that convergence is obtained with increasing values of $N$.}
In Fig. \ref{fig3}, we also report $\pi(j)$ %
 for $R=256$ and $N=256$, which is  {in}
excellent agreement with what reported in Fig. {\ref{fig1}, with overlapping confidence intervals. In the
following, we will refer to the thermodynamic limit $N\rightarrow \infty$,
where the discrete set of the Lyapunov exponents indices $j=1,\ldots,N$ is
replaced by the continuous variable $x\in[0,1]$.
\begin{figure}[t]
\includegraphics[width=1.0\columnwidth]{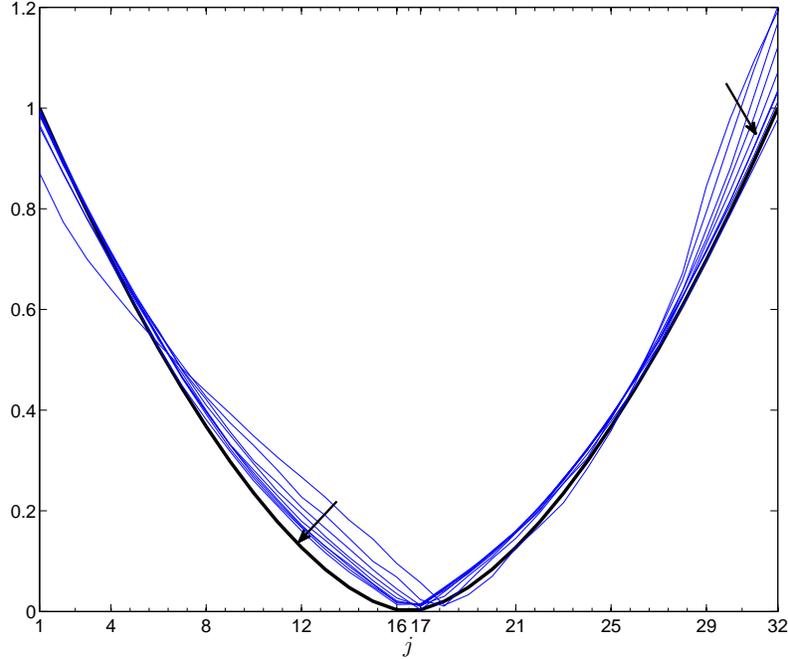}
\caption{Blue lines: $|\lambda_j+1|/(c_\lambda R^{2/3})$ for different $R$ (growing as indicated with the arrows from 8 to 2048 as powers of 2). Black line: $\left|2j/(N+1)-1\right|^{5/3}$. Irreversible model. See Eq. \ref{lexp}}
\label{fig5}
\end{figure}

Accurate scaling laws can be found also for the Lyapunov exponents, so that
it is possible to provide an extremely simple expression for the full
spectrum. We find stringent numerical evidence that
\begin{equation}
|\lambda(x)+\pi(x)| \sim c_\lambda \left|2x-1\right|^{5/3} R^{2/3}
\label{lexpexact}
\end{equation}
where $\pi(x)$ is depicted in Fig. \ref{fig2} or Fig. \ref{fig3} once we
interpret the index $j$ as $(N+1) x$. The scaling relation given in
Eq.(\ref{lexpexact}) can be safely approximated for large values of $R$ as:
\begin{equation}
|\lambda(x)+1| \sim c_\lambda \left|2x-1\right|^{5/3} R^{2/3}
\label{lexp}
\end{equation}
{\0{\it Remark:} In this case the largest Lyapunov exponent scales {\it
  independently} of $N$: a property that has been debated and is in
agreement with 
\cite{LPR986}.}
\*

In Fig. \ref{fig5} we provide a graphical evidence of Eq.(\ref{lexp}) {; we do not plot the confidence intervals of the rescaled $\lambda$'s as their width is comparable to or smaller than the width of the lines}. Note
that our results seem to disagree with \cite{Kar10}, where very different
empirical scaling laws are reported and tested on a much more limited range
of values of $R$ (up to about 30). While in the common range of $R$ the
results agree with what reported in \cite{Kar10}, we believe that the
scaling laws given here are more general because they provide an excellent
model for the obtained values of $\lambda$'s in a much wider range of
values of $R$. Moreover, it is clear that the expressions given in
\cite{Kar10} cannot be used for studying the limit of $R$ going to
infinity. The knowledge of $\lambda(x)$ allows one to compute several
relevant properties of the invariant measure of the system. We first find
the position $x_0$ of the vanishing Lyapunov exponent:
\begin{equation}
\lambda(x_0)=0 \rightarrow x=x_0 =\frac{1}{2}-\frac{R^{-2/3}}{2 c_\lambda^{3/5}}
\end{equation}
Obviously, if our system has $N$ degrees of freedom, we expect that
$\lambda_j=0$ {for} $j=\lceil Nx_0-0.5\rceil$. It is then
straightforward to derive the expression metric entropy, estimated as the
sum of the positive Lyapunov exponents \cite{ruelle89}:
\begin{equation}
\frac{\eta}{N}=\int_0^{x_0} dy \lambda(y) =\frac{3}{16}c_\lambda R^{2/3}-\frac{1}{2}+\frac{5}{16}R^{-2/5}c_\lambda^{-3/5},
\end{equation}
where the expression is valid in the limit of $N\rightarrow \infty$, and finally one can derive a closed expression for the Kaplan-Yorke dimension $d_{KY}$ defined as:
\begin{equation}
\frac{d_{KY}}{N}=\tilde{y}   {\qquad{\rm with}}
\qquad
 \int_0^{\tilde{y}} dy \lambda(y) =0
\end{equation}
In our case, since in our parametric range one can see that $\tilde{y}\geq 1/2$, we derive:
\begin{equation}
\int_0^{1/2} dy  \left(c_\lambda \left(1-2y\right)^{5/3} R^{2/3}-1\right) + \int_{1/2}^{\tilde{y}}dy \left(-c_\lambda \left(2y-1\right)^{5/3} R^{2/3}-1\right)=0
\end{equation}

We derive:
\begin{equation}
\int_{\tilde{y}}^{1} c_\lambda \left(2y-1\right)^{5/3} R^{2/3}= \tilde{y}  
\end{equation}
solving the integral we obtain:
\begin{equation}
\frac{3}{16} c_\lambda R^{2/3} \left( 1-\left(2\tilde{y}-1\right)^{8/3}\right) = \tilde{y}  
\end{equation}

This is an implicit formula for the Kaplan-Yorke dimension. Let's explore the limit where $\tilde{y}\sim 1$. Assuming $1-\tilde{y}\ll 1$ we derive:
%
%
\begin{equation}
 c_\lambda R^{2/3}  (1-\tilde{y}) = \tilde{y}  
\end{equation}

So that:

\begin{equation}
\tilde{y}  =1 -  \frac{1}{1+ c_\lambda R^{2/3} } 
\end{equation}

Therefore, if we have:
\begin{equation}
N-d_{KY} <1 \rightarrow  \frac{N}{1+ c_\lambda R^{2/3}} <1 
\end{equation}
the attractor has a dimension virtually indistinguishable from that of the
full phase space. {If $R$ is very large, the system's attractor 
occupies a phase space volume close to the one ergodically visited in
the inviscid, not forced system
or, equivalently, one could say the system is weakly damped}.  Note that
the last inequality can be read as $N<|\lambda_N|$, which agrees with what
would {be obtained} without resorting to the continuous variable $x$,
except for a term of order 1 ($N-1$ instead of $N$ in the \textit{l.h.s.}).

\def\SEC{Properties of the Irreversible Lorenz '96 Model}
\section{Equivalence between the Reversible and 
Irreversible Versions of the Lorenz '96 Model}\label{sec3}\setcounter{equation}{0}

Following the indications given in Eq.(\ref{e1.3A})-(\ref{e1.3C}), we
construct the candidate equivalent time-reversible model as follows:
\be\eqalign{ &\dot X_j=X_{j-1}(X_{j+1}-X_{j-2})+R-\a(X) X_j, \qquad
  j=0,\ldots,N-1 \cr &\a(X)=R\frac{\sum_j X_j}{\sum_j
    X_j^2}={\frac{RM}{2E}},\cr} \Eq{e1.6} \ee 
plus the imposed periodic conditions $X_k=X_{N+k}$ $\forall k$. The model
conserves exactly the energy $E=\sum_{j=1}^N X_j^2/2$. Following the
procedure described in Sec. \ref{sec1}, we initiate each simulation with a
given $R$ using initial conditions $X_j(0)$ $j=1,\ldots,N$, such that
$E(0)={\lis E^i_R}$, which makes sure that the energy of the flow realized
by the reversible model has constant energy equal to the average energy
observed in the corresponding irreversible model forced with the same value
of $R$.

Geometrically, the motion of the reversible model is constrained to
the spherical surface defined by the initial conditions and will reach a
stationary state defined by the measure $d\mu(x)^r_R$. In order to simulate
accurately such a delicately constrained dynamics it is extremely important
to use an integrator of ordinary differential equations able to use
adaptive step and allowing for pre-defined relative and absolute error in
the integration. Using the MATLAB\textregistered $ $V. 7.9 routine
\texttt{ode45} proved crucial for obtaining accurate results, because we
could keep the same standard of accuracy in the irreversible and reversible
runs. See Appendix \ref{scr1}. As a result of numerical errors, a very small amount of energy is
lost with time in the runs of the reversible model. Such spurious loss
increases with $R$ but we have managed to limit it in relative terms to
less than $10^{-6}$ with respect to the initial value between the beginning
and the end of the simulations.

Clearly, for a given value of $R$, the support of the invariant measures
$d\mu^i_R$ and $d\mu^r_R$ of the irreversible model (see Eq.(\ref{e1.2}))
and of the corresponding reversible model (see Eq.(\ref{e1.6})),
respectively, are substantially different, because $E$ in the irreversible
model fluctuates by $20-40\%$ (see Eq.(\ref{stdE})). Despite the fact that
the two attractors are clearly not coincident, the conjectured equivalence
holds to a very high degree of approximation. 

The first observation is that for all considered values of $R$ the
expectation value of $\a(X)$ is 1 within $1 \%$, in agreement with property
{\bf (1)} in Sec. \ref{equivalence}. As for point {\bf (2)}, we consider as
natural test observable $g(x)$ the quantity $M$. Also in this case, we
observe that $\lis M^i_R$ and $\lis M^r_R$ agree within $1 \%$, and the
applies same when comparing $\overline{M^2}^i_R$ and
$\overline{M^2}^r_R$. In Fig. \ref{fig5bis} we show the probability
distribution of $M$ for $R=512$ in order to give a feeling of the strong
similarity between the probability distribution functions (pdfs) of $M$ for
two long $R=256$ integrations performed using the reversible and
irreversible model.  Another finding is that, for the same $R$, the decorrelation time of $M$ in
the reversible and irreversible models agree accurately, so that also for the reversible model we can write $t_{dec,M}= c_{t,M} R^{-2/3}$.
\begin{figure}[ht]
\includegraphics[width=1.\columnwidth]{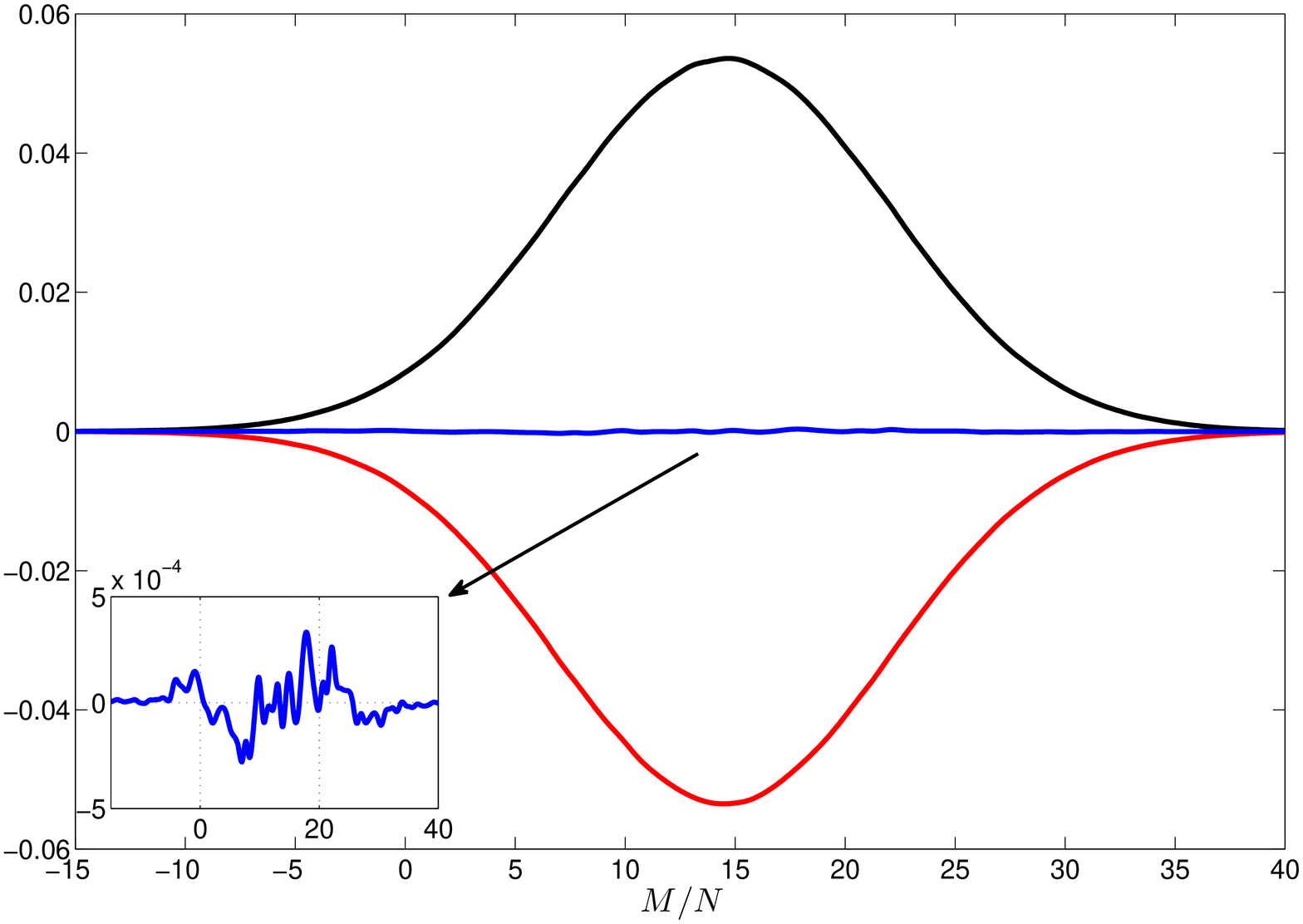}
\caption{Black line: pdf for $M/N$ in the reversible model, $R=2048$. Blue line: opposite of the pdf for $M/N$ in the irreversible mode, for the same value $R=2048$. Red line: sum of the black and blue line. Note the different vertical scale in the insert.}
\label{fig5bis}
\end{figure}

We now would like to address property {\bf (3)} in
Sec. \ref{equivalence}. The goal is to investigate the fluctuations at
different time scales of the quantity $\sigma(X)=(N-1)\alpha(x)$. {For this
purpose, at a given value of $R$,} we define: \be
p^x_{\tau,R}=\frac1\t\int_0^\t \frac{\s(S_t x)}{\lis\s_R} \ee 
as the
$\tau-$averaged value of the contraction of the phase space evaluated
starting from point $x$ {
initially chosen at random uniformly on the energy surface
and, therefore, with statistical properties that with probability $1$
coincide with those of a typical point on the attractor,
described by the invariant SRB measure of the system
defined by the parameter $R$}, and consider that $\lis\s_R\sim N-1$ within
$1\%$, $\forall R$. Let's define $P^R_\tau(p)dp$ the probability
(constructed according to the invariant measure of the system) of observing
$p^x_{\tau,R}\in [p,p+dp]$. We want to test whether the FR holds, namely,
whether:
\begin{equation}
\frac{1}{\tau}\log\left[\frac{P^R_\tau(p)}{P^R_\tau(-p)}\right]=\overline\sigma_R p +O(\tau^{-1}),
\label{frg}\end{equation}
and we want to study also {which} is the rate of convergence of the FR
to its ${\tau \rightarrow \infty}$ limit, in order to substantiate the
extent to which the FR can be verified, and be useful, when finite times
(in the macroscopic time scales of the system) are considered.

We construct from the original time series of $\sigma(t)$ given as output
of the reversible model the corresponding $\tau-$coarse grained time
series. We then construct the $P_\tau$ probability distributions by either
using non-parametric estimators such as histograms (tested for numerical
stability) or, instead, Gaussian kernel smoothers \cite{Bo97}. 

The results
presented here are weakly dependent on the statistical model used for
estimating probability density,  { but using histograms it is easier to construct confidence intervals for the left hand side of Eq. \ref{frg}. Therefore,  we present findings obtained with histograms}. We have that, since {$\sigma=\frac{RM}{2E}$}, 
$\sigma/\overline\sigma_R$
and $M/\overline{M}^r_R$ are identical, so that,
$t_{dec,\sigma}^{R}=t_{dec,M}^{R}$, where we have added explicit reference
to $R$. From what discussed in Sec. \ref{sec2}, it is more likely to find
negative fluctuations for the $\tau$-time averaged values of $\sigma$ when
high values of $R$ are considered. We present results relative to $R=512$
and $R=2048$ in Figs. \ref{fig6}-\ref{fig7}, respectively.  We plot:
\begin{equation}
\frac{1}{\tau\overline\sigma_R}\log\left[\frac{P^R_\tau(p)}{P^R_\tau(-p)} \right]
\label{frgplot}\end{equation}
against $p$.  {The  $3\sigma$ confidence intervals are computed by suitably performing  bootstrap on the results obtained on steady state trajectories of length $ 10^6 t_{dec,\sigma}^R$.  } Three main observations can be made:
\begin{itemize}
\item For both values of $R$, the FR is obeyed to a good degree of accuracy
  for \textit{large} - to be defined later - values of $\tau$;
\item Before the asymptotic result is obtained, the quantity given in
  Eq. (\ref{frgplot}) is proportional to $p$ via a constant $c(\tau)$, which
  converges to $1$ as $\tau$ becomes large;
\item The larger $R$, the faster is the convergence to the FR with $\tau$.
\end{itemize}
\begin{figure}[ht]
\includegraphics[trim=0cm 0cm 0cm 0 cm, clip=true, width=1.\columnwidth]{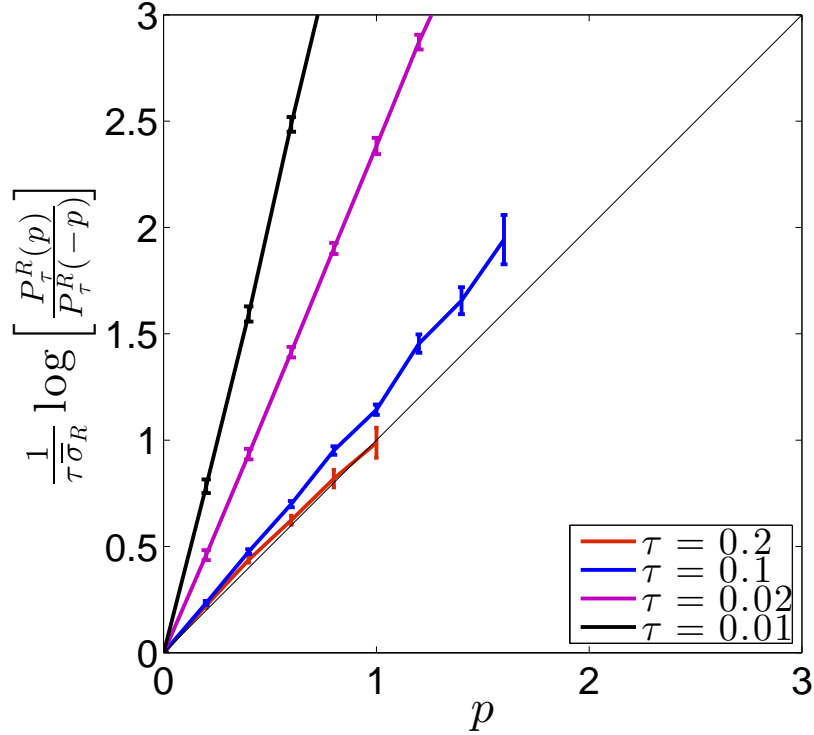}
\caption{Fluctuation Relation for $R=512$, {approaching slope $1$ as $\t$
  increases}. See Eq. \ref{frg}.}
\label{fig6}
\end{figure}

\begin{figure}[ht]
\includegraphics[trim=0cm 0cm 0cm 0cm, clip=true, width=1.\columnwidth]{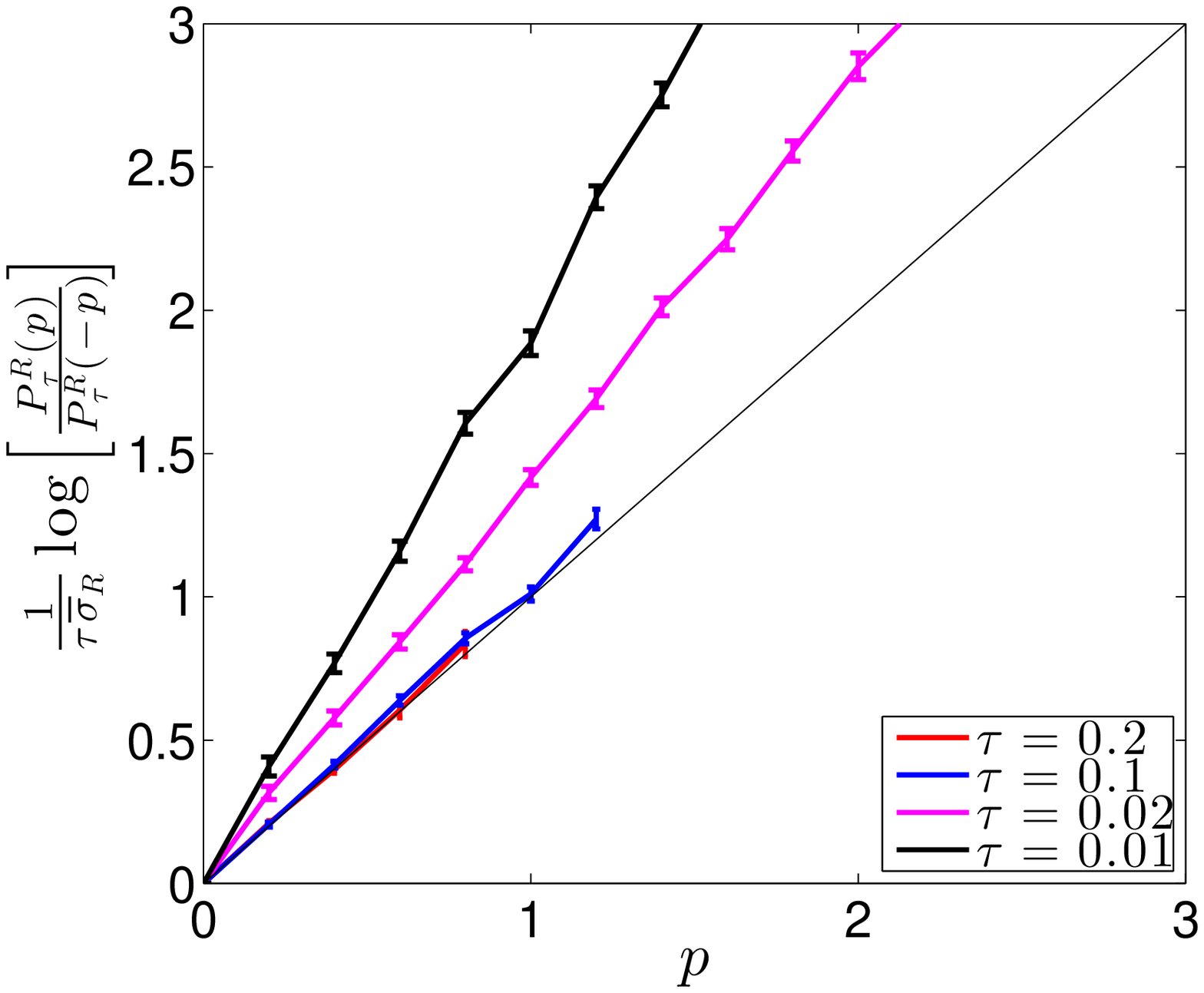}
\caption{Fluctuation Relation for $R=2048$, {with slope approaching $1$
    as $\t$ increases beyond the decorrelation time}. See
  Eq.(\ref{frg}). Note that the red line is hidden below the blue line. }
\label{fig7}
\end{figure}

The first observations basically provides a verification for point {\bf
  (3)} of the conjecture proposed in Sec. \ref{equivalence}.  The second
observation does not derive trivially from the FR given in Eq.(\ref{frg}),
because it implies that the $O(\tau^{-1})$ correction is, in fact, proportional
to $p$. Since $p$ is of order one, this result is anyway consistent with
the FR.  {One can show that  $O(\tau^{-1})\propto p$ in the case of gaussian process, which, in fact, $P_\tau^R$ conforms to with a high degree of accuracy}. It is possible to infer from numerical experiments a very simple
expression for the corrective factor $c(\tau)$: 
\begin{equation}
c(\tau)=1+\left(\frac{t_{dec,\sigma}^{R}}{\tau}\right)^{4/3}=1+\left(\frac{c_{t,\sigma }}{\tau}\right)^{4/3}R^{-8/9},\label{correction}
\end{equation}
where $c_{t,\sigma }=c_{t,M }$, which explains the third observation above. Equation (\ref{correction}) is
crucial for understanding the speed of convergence of the limit implied by
the FR, and, conversely, for estimating how large $\tau$ must be in order
to find good agreement with the FR. Interestingly, the time scale appearing
in the formula is the decorrelation time of $\sigma$ itself, which is
extremely easy to estimate {and which we have observed to coincide with
  that of $M$}. Roughly, as long as $\tau$ is larger than such time scale
by a factor of say 10, we observe small deviations from the FR. Moreover,
the corrective factor convergences to 1 with $R^{-8/9}$, 
indicating that as the forcing to the system becomes more intense, the FR
relation can be observed for smaller and smaller averaging times $\tau$,
going to 0 in the limit $R\rightarrow \infty$.


The FR is indeed signature of the a) time-reversible nature of the flow and
b) of the compatibility with the chaotic hypothesis. Given the
equivalence discussed above, it makes sense to check whether in the
irreversible system one can find quantities obeying the FR. 

We construct as obvious candidate is
$\tilde{\sigma}_R=NRM/(2\overline{E}^i_R) $. The first moments of this
quantity, as discussed above referring to $M$, are almost identical to
those of ${\sigma}_R$ in the reversible system. Nonetheless, when testing
{whether the FR holds even in this irreversible case}, we find
deviations with respect to what obtained in the reversible case. In
particular, we find that $c_\tau$ does not converge to 1, with
discrepancies of the order of $20\%-30\%$. The presence of such - {not
too large} -
departure from the FR might explain why some experimental and numerical
attempts at verifying FR in irreversible systems, and so outside its realm
of validity, have encountered {some} degree of success.

Following a suggestion first discussed in {\cite{Ga000b,Ga006d}}, we
have also attempted the verification of a local version of the FR, which
amounts to studying the large fluctuations of some measure of the local
contraction of the phase space. Therefore, we define
{(\cite[Eq.(6.2)]{Ga000b},\cite[Eq.(9.2)]{Ga000a})}

\be \sigma^{\beta}(X)=N
R\frac{\sum_{j=1}^{N_0} X_j}{\sum_{j=1}^{N} X_j^2}.  \ee 
where the summation in the numerator is performed over $N_0$ nearby points, covering a
portion $\beta=N_0/N$ of the total volume of the system (note that the
system is invariant to discrete translation). In \cite{Ga000b} it is
{proposed} that\\ 
{\bf(a)} the expectation value of
$\overline{\sigma^\beta}_R=\beta\overline{\sigma}_R$; and\\
{\bf (b)} the
quantity {$p=\frac{\sigma^{\beta}}{\lis \sigma^\beta}$} obeys FR, so that,
defining as usual $p^x_{\tau,R}=\frac1\t\int_0^\t \frac{\s^\beta(S_t
  x)}{\lis\s_R}$, and $P^R_\tau(p)dp$ the probability of observing
$p^x_{\tau,R}\in [p,p+dp]$. we should have:
\begin{equation}
\frac{1}{\tau}\log\left[\frac{P^R_\tau(p)}{P^R_\tau(-p)}\right]=\overline{\sigma}^\beta_{R} p +O(\tau^{-1})=\beta\overline{\sigma}_{R} p +O(\tau^{-1})
\label{frgloc}\end{equation}
We have tested these predictions for $R=2048$ by considering $N=32$ and
$N_0=8$.   

%
%
\begin{figure}[ht]
  \includegraphics[trim=0cm 0cm 0cm 0cm, clip=true, width=1.\columnwidth]{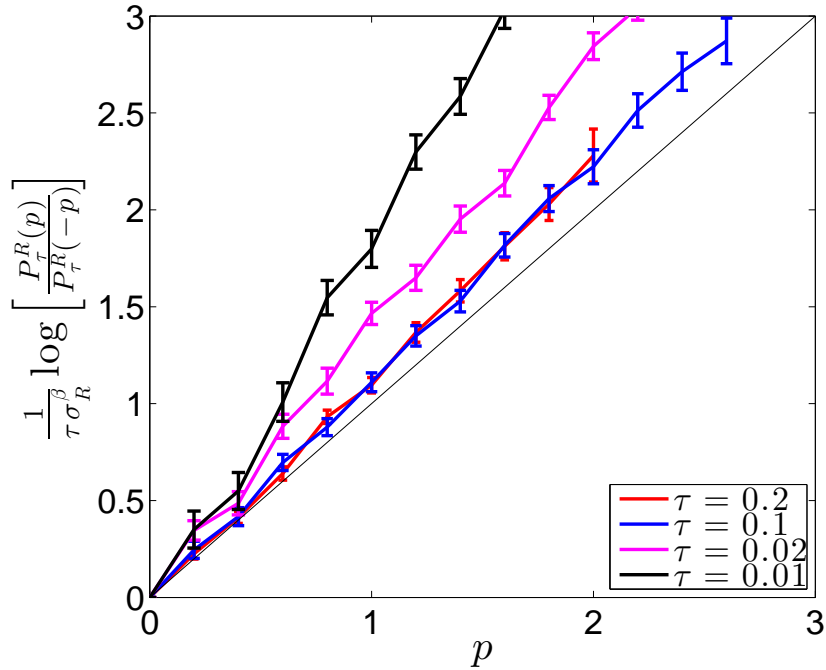}
\caption{Local version of the Fluctuation Relation for $R=2048$. See
  Eq. \ref{frgloc}. Note that the red line is hidden below the blue line. }
\label{fig7bis}
\end{figure}
\begin{figure}[ht]
\includegraphics[width=1.\columnwidth]{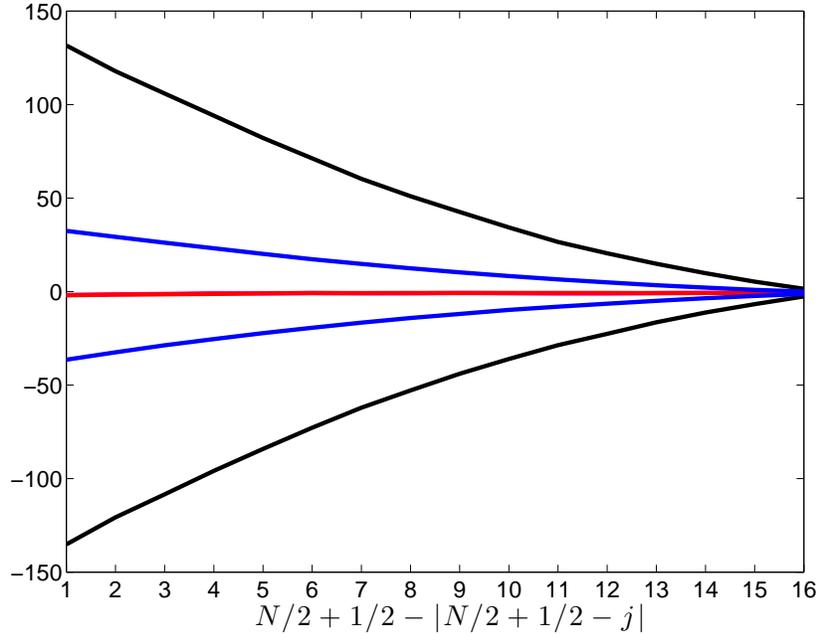}
\caption{Black line: Lyapunov exponents for $R=2048$ arranged pairwise. Magenta line: corresponding value of $(\lambda_j+\lambda_{N-j+1})/2$. Blue line: Lyapunov exponents for $R=256$ arranged pairwise. Red line: corresponding value of $(\lambda_j+\lambda_{N-j+1})/2$. Reversible model (Compare with Fig. \ref{fig1}).}
\label{fig8}
\end{figure}
\begin{figure}[ht]
\includegraphics[width=1.\columnwidth]{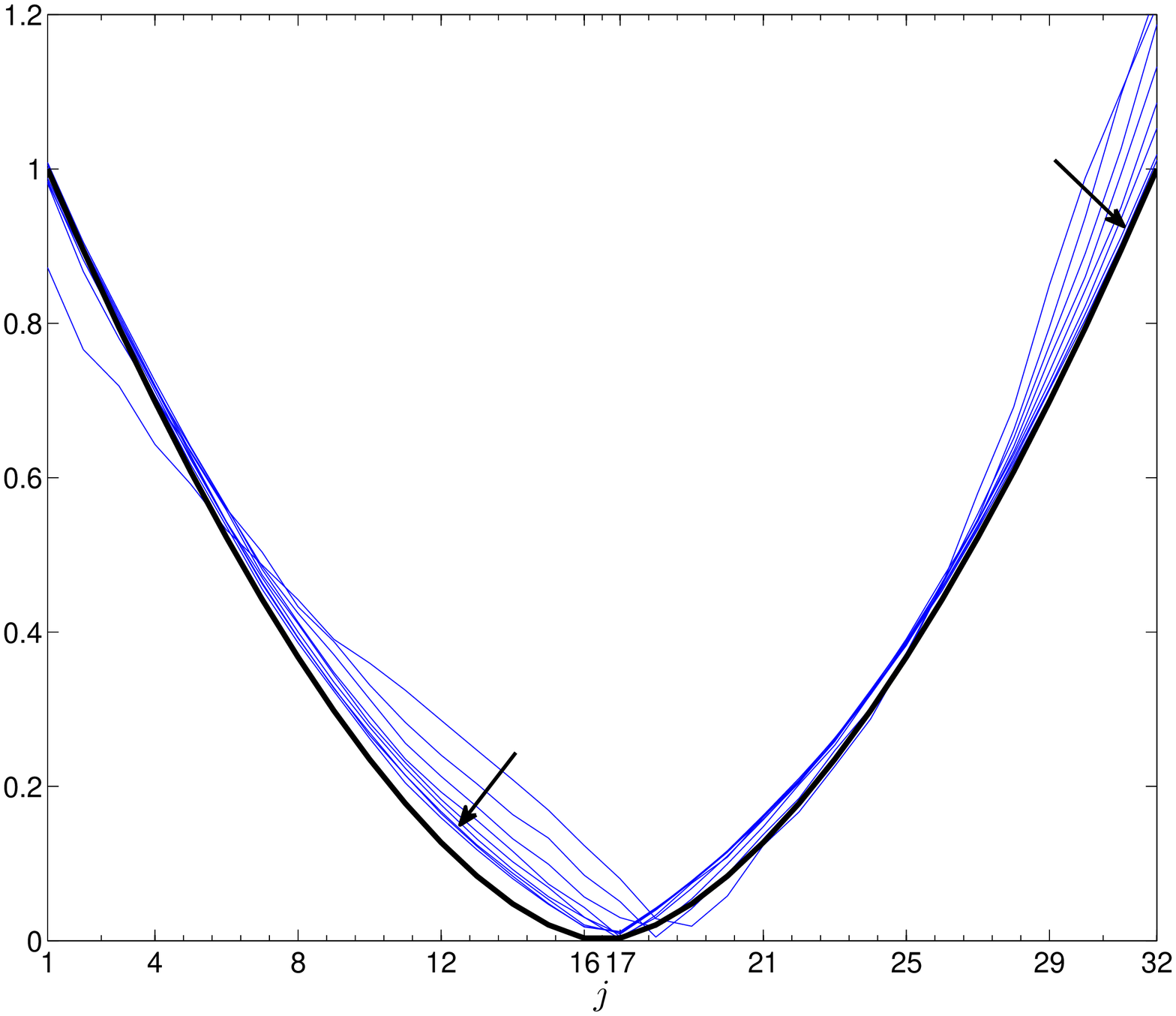}
\caption{Blue lines: $|\lambda_j+1|/(c_\lambda F^{2/3})$ for different $F$ (growing as indicated with the arrows from 8 to 2048 as powers of 2). Black line: $\left|2j/(N+1)-1\right|^{5/3}$. Reversible Model. See Eq. \ref{lexp} and compare with Fig. \ref{fig3}.}
\label{fig9}
\end{figure}

The verification of property {\bf(a)} is rather easy.   {Figure \ref{fig7bis} shows that the local version of the FR proposed as  property {\bf (b)}  is not perfectly obeyed (we observe a deviation of 5-10\%), possibly because of a finite size effect, because in our experimental conditions we are far from the limits proposed in the conjecture above $\beta\rightarrow 0$, $N_0, N\rightarrow\infty$}. 
  
As for addressing point {\bf (4)} in the list of properties discussed in
Sec. \ref{equivalence}, we compute the spectrum of the Lyapunov exponents
of the reversible model given in Eq. \ref{e1.6}. See Appendix
\ref{scr2}. It is important to note that, given the integral nature of the
fluctuating viscosity $\alpha(X)$, the Jacobian matrix to be used in the
Benettin et al. \cite{benettin1980lyapunov} algorithm is not anymore
sparse, with resulting increased computational costs in terms of memory.

Our result provide strong evidence that the spectrum of Lyapunov exponents
in the reversible and irreversible cases are indeed the same within statistical 
uncertainty  {(\textit{i.e.} error bars overlap)}. In Fig. \ref{fig8} we present the Lyapunov
spectrum obtained for $R=256$ and $R=2048$ and $N=32$, which, by visual
inspection, have a very close correspondence to what shown in
Fig. \ref{fig1}.  {Error bars are not reported in Fig. \ref{fig8} to improve readability.}


The quality of the agreement is clarified by the fact that also in the
reversible case the approximate scaling law $\lambda(x)=c_\lambda
(2x-1)^{5/3}R^{2/3}$ applies; compare Fig. \ref{fig9} with
Fig. \ref{fig3}. The accuracy of the correspondence between the Lyapunov
{exponents} of the irreversible and reversible case is further substantiated by {the fact that we find virtually the same pairing rule $\pi(x)$ for all values of R (not shown) }
Since the
scaling of the Lyapunov vectors agrees, and the pairing rule, which is
obtained as the difference between two large numbers, is the same when
comparing the irreversible and reversible case, we are drawn to the
conclusion that also property {\bf (4)} is indeed accurately obeyed.


\def\SEC{Summary and Conclusions}
\section{Summary and Conclusions}\label{sec4}\setcounter{equation}{0}

The possibility of providing macroscopically equivalent yet microscopically
non identical descriptions of many-particle systems endows equilibrium
statistical mechanics with a great flexibility in terms of theoretical
calculations, and leads directly to a robust definition of thermodynamical
properties. In this paper we have examined the problem of constructing
equivalent ensembles for non-equilibrium statistical mechanical systems.

We have approached this problem by treating {a} case of irreversible
dissipative balanced systems, {a class of systems which includes the NS
equations}, characterized by a set of friction coefficients $\Bn$ and a
corresponding set of quantities $\V E$ which are conserved if forcing and
dissipation are neglected, and fluctuate in time when forcing and
dissipation are considered and the system reaches its invariant measure
$\m^i_{\Bn}$.

 The corresponding reversible systems are constructed by changing the
 definition of constant friction into functions $\Ba(x)$ in such a way that
 the quantities $\V E$ are conserved (and set to the average values
 $\lis{\V E}^i(\Bn)$ of the functions $\V E(x)$ in the distribution
 $\m^i_\Bn$). 

The attractor of such a system is the support of the invariant
 measure $\m^r_{\lis{\V E}}$. While the two invariant measures are indeed
 different, we can say that $\m^i_\Bn$ corresponds to $\m^r_{\V E}$ if
 $\m^r_{\V E}(\Ba)=\Bn$. 
 We have proposed four criteria for defining the
 equivalence of the irreversible and reversible dynamics, which boils down
 to providing analogous information of the expectation values of smooth
 observables, to featuring an equivalent description of the instabilities
 of the system, and, in the spirit of the chaotic hypothesis, to featuring,
 in the case of the reversible system, a phase space contraction rate
 fluctuating in agreement with the FR.
    
    While the long-term goal of this investigation is \textit{e.g.} to
    study the equivalence between the customary irreversible representation
    of fluid-dynamics through the NS equations and its reversible version,
    in this paper we have considered a simple yet dynamically rich model
    introduced by E. Lorenz \cite{LE998,Lo005}. Such a model describes
    advection, dissipation, forcing, in a ring of $N$ grid points, and has
    been the subject of extensive investigation in atmospheric sciences (see, e.g. \cite{Orr03,Wil05,Tre04,Tre10} ) 
     and, more recently, in statistical physics (see, e.g. \cite{majda07,Hal10,lucarini2011}). This
    model fits the paradigm of the irreversible dissipative balanced
    systems, so that, when forcing (described by the parameter $R$) and
    dissipation (a normalized viscosity $\nu$ is set to 1) are neglected, a
    quadratic quantity referred to as energy $E$ is conserved.

The Lorenz '96 system is strongly chaotic when the forcing parameter $R$ is
larger than 8 and its properties are extensive with the number of nodes
$N$.  Our investigations have covered extensively the range $R=8$ up to
$R=2048$, plus control runs performed at much
higher value of $R$. Most of the simulations have been performed
considering $N=32$, plus several additional runs performed with $N=256$ and
$N=512$ to check how properties scale with $N$.

We have confirmed the existence of accurate scaling laws allowing to
express the energy $E\sim R^{4/3}$ of the system and a suitably linear
quantity referred to as momentum $M\sim R^{1/3} $. More interestingly, it
is possible to find evidence of a simple scaling relation for the Lyapunov
vectors $\lambda$'s such that $|\lambda(x)+\pi(x)|\sim(2x-1)^{5/3}R^{2/3}$,
where $x$ refers to the normalized index $j/(N+1)$ in the continuous limit, and
$\pi(x)$ is a {$R$ independent pairing function, such that
  $\lambda(x)+\lambda(1-x)=2\pi(x)$ $\forall x\in [0,1]$ and (at least) for
  all the $R$'s considered}.

We have found {numerical evidence for} explicit expressions of the
position of the zero Lyapunov exponent, of the metric entropy, and of the
Kaplan-Yorke dimension. The Kaplan-Yorke dimension saturates the dimension
of the phase space as $R\rightarrow\infty$, as already observed by Lorenz
\cite{Lo005}, {although the statistical properties, both in the
  reversible and in the irreversible models, are very different from those
  of the uniform distribution on the energy surface, \ie from the
  equilibrium statistics in spite of the choice of the initial condition
  which are selected precisely with the latter distribution: a property
  familiar, from SRB theory of chaotic motions, in SRB distributions}.

In the irreversible model $E$ fluctuates, while the contraction of the phase
space is fixed. We can think this as analogous to the canonical ensemble,
where the viscosity plays the role of temperature, while $E$ plays the role
of energy.
    
Following the paradigm described above, we have constructed the (candidate)
equivalent reversible system by introducing a fluctuating viscosity
allowing for energy conservation and defining a time-reversible
dynamics. Therefore, in this model the phase contraction fluctuates and
energy {is} fixed, as opposed to the traditional irreversible model. We
have verified for $\forall R$ excellent agreement between the reversible
and irreversible model on {\bf (1)} the average phase contraction rate; and
{\bf (2)} first moments of macroscopic observables (here we have taken $M$
as obvious candidate).

We have also verified {\bf (3)} that the FR for the phase space contraction
rate $\sigma_R$ is obeyed with good accuracy, and have highlighted that the
convergence to the asymptotic result is controlled by the ratio between the
averaging time $\tau$ over which the phase space contraction is evaluated
and its $e-$folding time. 

One finds that as long as $\tau$ is larger by a
factor of say $5-10$, the FR is obeyed to a very good approximation. For
any finite value of $\tau$, the correction to the FR result $\rightarrow
0\sim R^{-8/9}$, so that for strong forcings the FR can be observed over
very short averaging times. Interestingly, the finite-time correction to
the asymptotic FR result appears as a multiplicative tending to 1 rather than
additive factor going to 0 with $\tau\rightarrow\infty$, and can be
related, in the Gaussian case, to the scaling of the variance of the
distribution of $\tau-$averaged values of $\sigma_R$. We have also for the
first time verified the validity of a local version of the FR proposed in
\cite{Ga000b}  {within a good degree of approximation.}

Finally, we have also compared \textbf{(4)} the instability properties of
the irreversible and reversible system by investigating the degree of
agreement of the spectrum of Lyapunov exponents. We find that an extremely
satisfying agreement, so that the same scaling laws and pairing rule
{are} obeyed. When $R$ is very large, the system lives on an attractor
whose closure is indistinguishable from the spherical shell defined by the
initial value of the energy, thus resembling a micro-canonical ensemble.
 
We conclude by stressing that while the support of the two invariant measures
$\mu^i_R$ and $\mu^r_R$ is indeed different, and the symmetries of the two
corresponding dynamical systems also are different, {numerical evidence
  supports, in the Lorenz96 system, that} their statistical mechanical
properties are virtually the same at all considered values of $R$ where
turbulence is observed. For high $R$, we basically find an analog of the
  equivalence of canonical and microcanonical ensembles in chaotic
  systems.

{In other words} one may naively think that for
high values of $R$ the constructed time-reversible system may, in fact,
feature the same statistical mechanical properties of the system where
forcing and dissipation are neglected altogether, if they live on the same
energy shell. Note that value of $\overline{E}^r_R$ and $R$ are one-to-one
related. This is, in fact, false. The first remark is that while the
expectation value $\overline{M}^r_R\sim R^{1/3}$ in the time reversible
system, in the corresponding inviscid and unforced system the expectation
value of $M$ is, by symmetry, zero. Moreover, as discussed before, the
Lyapunov exponents of the inviscid and unforced dynamics and of the
time-reversible dynamics are indeed different (even if the difference is
smaller and smaller in relative terms as larger and larger energies are
considered). Note that the empirically defined pairing rule convergences
for large values of $R$ to $\pi(x)$ which is indeed
non-vanishing. Therefore, the SRB measure is distinct from the Gibbs
measure \textit{also in the limit of $R\rightarrow\infty$}, even though the
attracting sets of the two measures become indistinguishable, with the sign
of $R$ responsible for breaking the symmetry.

These results provide a strong motivation for considering the possibility
of approaching the problem of modeling turbulent fluids using a radically
different point of view, where, instead of setting to a constant the rate
of contraction of the phase space, we keep constant some physical
quantities of interest. This approach is expected to work more effectively
when the nondimensional numbers describing the relative strength of forcing
versus dissipation (e.g. the Reynolds number) is very large. One can think
of applying this procedure only to specific regions of a fluid, {as
  suggested by the local result in Fig. \ref{fig7bis}}. 

This seems promising in the case
we have a fluid with strong anisotropy in one direction, like in the case
of strong stratification, of special interest for geophysical
applications. Moreover, these results clarify the relevance of the Chaotic
Hypothesis for investigating turbulent fluid systems. In particular, under
the assumption that the Chaotic Hypothesis holds, one can use the Ruelle
response theory \cite{ruelle98,ruelle2009} for studying the response of a
system like the climate to forcings (see discussions in \cite{lucarini2008,lucarini2009} and some results in
\cite{lucarini2011,Luc2014,Rag2014}).

As next steps in the line of investigation of this paper, {one can
  envisage to} approach the problem of studying the quasi-geostrophic
turbulence in a realistic Earth-like setting, trying to go from a point of
view where diffusion, Eckman friction, and viscosity are kept constant, to
a point of view where the total energy and potential enstrophy at various
levels are kept constant.
\newpage
   \appendix
\small
\def\SEC{Script - Part 1}
   \section{Script - Part 1: Integrating the Lorenz 96 irreversible and reversible models}\label{scr1}
   \begin{verbatim}
   %%%%%%%%%%%%%%

%%%% PART 1: Integrating the equations for the Loren6 96 
%%%% irreversible and reversible models
%%%%%%%%%% Script by V. Lucarini (valerio.lucarini@uni-hamburg.de)
%%%%%%%%%% Can be freely distributed according to GNU license. 

%%%%%%%%% Companion Material to the paper
%%%%%%%%% Equivalence of Non-equilibrium Ensembles, 
%%%%%%%%% and Representation of Friction in Turbulent Flows: the 
%%%%%%%%% Lorenz 96 model, by G. Gallavotti and V. Lucarini, 2014
%%%%%%%%% Values given below for the parameters 
%%%%%%%%% must be checked/optimized by the user

%%%%%%%%%%%%
% Lorenz 96 model - Irreversible version. Input parameters: 
% N = number of modes ; R = R parameter
%%%%%%%%%%%%%%%%%%%%%%%%%%%%%%%%%%%%%%%%%%%%%%%%%%%%%%%%%%%%%%
%%%save text below as lorenz96irr.m%%%%%%%%%%%%%%%%%%%%%%%%%%%
function dy = lorenz96irr(t,y,R,N)
dy=zeros(N,1);
F=ones(N,1)*R;
dy(1)=(y(2)-y(N-1))*y(N)-y(1)+F(1);
dy(2)=(y(3)-y(N))*y(1)-y(2)+F(2);
dy(N)=(y(1)-y(N-2))*y(N-1)-y(N)+F(N);
for j=3:N-1
dy(j)=(y(j+1)-y(j-2))*y(j-1)-y(j)+F(j);
end
%%%%%%%%%%%%%%%%%%%%%%%%%%%%%%%%%%%%%%%%%%%%%%%%%%%%%%%%%%%%%%
%%%save text above as lorenz96irr.m%%%%%%%%%%%%%%%%%%%%%%%%%%%

%%%%%%%%%%%%%%%%%%%%%%%%%%%%%%%%%%%%%%%%%%%%%%%%%%%%%%%%%%%%%%

odeset 
% Defines options for ode integrator (we use the standard 
% ode45 function)

%           AbsTol: [ positive scalar or vector {1e-6} ]
%           RelTol: [ positive scalar {1e-3} ]
%      NormControl: [ on | {off} ]
%      NonNegative: [ vector of integers ]
%        OutputFcn: [ function_handle ]
%        OutputSel: [ vector of integers ]
%           Refine: [ positive integer ]
%            Stats: [ on | {off} ]
%      InitialStep: [ positive scalar ]
%          MaxStep: [ positive scalar ]
%              BDF: [ on | {off} ]
%         MaxOrder: [ 1 | 2 | 3 | 4 | {5} ]
%         Nacobian: [ matrix | function_handle ]
%         JPattern: [ sparse matrix ]
%       Vectorized: [ on | {off} ]
%             Mass: [ matrix | function_handle ]
% MStateDependence: [ none | {weak} | strong ]
%        MvPattern: [ sparse matrix ]
%     MassSingular: [ yes | no | {maybe} ]
%     InitialSlope: [ vector ]
%           Events: [ function_handle ]

options=odeset('RelTol',1e-8,'AbsTol',1e-8);
% Our standard choice; a very good precision is enforced

N = 32;
% We consider the case of N = 32

F=2.^[3:11]; 
% Here is a vector of values of the forcing R

meanMirr=zeros(size(F)); 
% Initialization vector computing the mean values of M

meanEirr=zeros(size(F)); 
% Initialization vector computing the mean values of E

T=100; 
% Final time

dt=0.001; 
% Output is given at each dt

for j=1:max(size(F)); 
    % For loop 

    j   
    % to keep track of things
    
ci=randn(N,1); 
% Random initial condition

tic; 
%Start clock

[Tirr,Yirr]=ode45(@lorenz96irr,[0:dt:T-dt],ci,options,F(j),N); 
% Integrates the function lorenz96irr above (to be saved in a 
% separate file as lorenz96irr.m )

Yirr=Yirr(end/10:end,1:end); 
% let's remove some initial transient behavior

toc 
% Says elapsed time; just to check how long the integration is

Mirr=sum(Yirr,2)/N; 
% Momentum M (divided by N)

Eirr=sum(Yirr.^2/2,2)/N; % Energy (divided by N)

meanMirr(j)=mean(Mirr); % mean Momentum
meanEirr(j)=mean(Eirr); % mean Energy 

end;  
% end for loop; we have the statistics of the 
% Lorenz 96 irreversible model.

%%%%%%%%%%%%%%% Now to the reversible case

% Lorenz 96 model - Reversible version. 
% Input parameters: N = number of modes ; R = R parameter

%%%%%%%%%%%%%%%%%%%%%%%%%%%%%%%%%%%%%%%%%%%%%%%%%%%%%%%%%
%%%save text below as lorenz96rev.m%%%%%%%%%%%%%%%%%%%%%%
function dy = lorenz96rev(t,y,R,N)
dy=zeros(N,1);
F=ones(N,1)*R;
Fg2=ones(N,1)*f*sum(y)/sum(y.^2); % Note: this is the 
                                  % fluctuating viscosity 
dy(1)=(y(2)-y(N-1))*y(N)-Fg2(1)*y(1)+F(1);
dy(2)=(y(3)-y(N))*y(1)-Fg2(2)*y(2)+F(2);
dy(N)=(y(1)-y(N-2))*y(N-1)-Fg2(N)*y(N)+F(N);
for j=3:N-1
dy(j)=(y(j+1)-y(j-2))*y(j-1)-Fg2(j)*y(j)+F(j);
end
%%%%%%%%%%%%%%%%%%%%%%%%%%%%%%%%%%%%%%%%%%%%%%%%%%%%%%%%%%
%%%save text above as lorenz96rev.m%%%%%%%%%%%%%%%%%%%%%%%
%%%%%%%%%%%%%%%%%%%%%%%%%%%%%%%%%%%%%%%%%%%%%%%%%%%%%%%%%%

% Now we start with the same loop as above
options=odeset('RelTol',1e-8,'AbsTol',1e-8);

N =32;
% We consider the case of N = 32

F=2.^[3:11]; 
% Here is a vector of values of the forcing R

meanMrev=zeros(size(F)); 
% Initialization vector computing the mean values of M

meanErev=zeros(size(F)); 
% Initialization vector computing the mean values of E

T=1; 
% Final time

dt=0.001 
% Output is given at each dt

for j=1:max(size(F)); 
    % For loop 

    j 
    % to keep track of things
    
ci=randn(N,1); 
% Random initial condition: Step 1

nor=sum(ci.^2/2/N); ci=ci*sqrt(meanEirr(j))/sqrt(nor); 
% Random initial condition: Step 2: the system is 
% initializated from the energy shell E=meanEirr(j)

tic; 
%Start clock

[Trev,Yrev]=ode45(@lorenz96rev,[0:dt:T-dt],ci,options,F(j),N); 
% Integrates the function lorenz96rev above (to be saved in a  
% separate file as lorenz96rev.m )

Yrev=Yrev(end/10:end,1:end); 
% let's remove some initial transient behavior; not necessary

toc 
% Says elapsed time; just to check how long the integration is

Mrev=sum(Yrev,2)/N; 
% Momentum M (divided by N)

Erev=sum(Yrev.^2/2,2)/N; 
% Energy (divided by N): note: the energy is constant 
% to a very high degree of accuracy (controlled by options)

meanMrev(j)=mean(Mrev); 
% mean Momentum

meanErev(j)=mean(Erev); 
% mean Momentum 

end;  
% end for loop; we have the statistics of the 
% Lorenz 96 reversible model. Go to Part 2 in Appendix B
\end{verbatim}
\newpage

\def\SEC{Script - Part 2}
\section{Script - Part 2: Computing the Lyapunov exponents for the Lorenz 96 irreversible and reversible models}\label{scr2} 
\begin{verbatim}
%%%%%%%%%%%%%%

%%%% PART 2: Computing the Lyapunov Exponents from the irreversible
%%%% and reversible models
%%%% 
%%%%%%%%%% Script by V. Lucarini (valerio.lucarini@uni-hamburg.de)
%%%%%%%%%% Can be freely distributed according to GNU license. 

%%%%%%%%% Companion Material to the paper
%%%%%%%%% Equivalence of Non-equilibrium Ensembles, and 
%%%%%%%%% Representation of Friction in Turbulent Flows: 
%%%%%%%%% the Lorenz 96 model, by G.
%%%%%%%%% Gallavotti and V. Lucarini, 2014
%%%%%%%%% Values given below for the parameters 
%%%%%%%%% must be checked/optimized by the user

% Lorenz 96 extended model (trajectory and Jacobian) - Irreversible 
% version. Input parameters: N = numbers of modes ; R = R parameter
%%%%%%%%%%%%%%%%%%%%%%%%%%%%%%%%%%%%%%%%%%%%%%%%%%%%%%%%%%%%%%%%%%%
%%%save text below as lorenz_ext96irr.m%%%%%%%%%%%%%%%%%%%%%%%%%%%%
function dy=lorenz_ext96irr(t,y,R,N)

%%%%%%%%%  Same as lorenz96irr

dy = zeros(N*(N+1),1);    % a column vector
F=ones(N,1)*R;
dy(1)=(y(2)-y(N-1))*y(N)-y(1)+F(1);
dy(2)=(y(3)-y(N))*y(1)-y(2)+F(2);
dy(N)=(y(1)-y(N-2))*y(N-1)-y(N)+F(N);
for j=3:N-1
dy(j)=(y(j+1)-y(j-2))*y(j-1)-y(j)+F(j);
end

%%% Initializing the Jacobian 

Y=zeros(N,N);

for j=1:N;
    for k=1:N;
        Y(j,k)=y(N+j+(k-1)*N);
    end;
end;

% Linearized system
% We split the jacobian in two parts

Jac1=zeros(N,N); 
Jac2=zeros(N,N);

Jac1=-eye(N); % Frictional part

% Now the rest, 

Jac2(1,2)=y(N);
Jac2(1,N)=y(2);
Jac2(1,N-1)=-y(N);

Jac2(2,3)=y(1);
Jac2(2,N)=-y(1);
Jac2(2,1)=y(3)-y(N);

Jac2(N,1)=y(N-1);
Jac2(N,N-2)=-y(N-1);
Jac2(N,N-1)=y(1)-y(N-2);

for j=3:N-1
Jac2(j,j+1)=y(j-1);
Jac2(j,j-1)=y(j+1)-y(j-2);
Jac2(j,j-2)=-y(j-1);
end

Jac=Jac1+Jac2; %the Jacobian is now obtained

%Variational equation   
dy(N+1:N*(N+1))=Jac*Y; 
%Note: MATLAB rearranges correctly the matrix into a vector 
%%%%%%%%%%%%%%%%%%%%%%%%%%%%%%%%%%%%%%%%%%%%%%%%%%%%%%%%%%%%%
%%%save text above as lorenz_ext96irr.m%%%%%%%%%%%%%%%%%%%%%%
%%%%%%%%%%%%%%%%%%%%%%%%%%%%%%%%%%%%%%%%%%%%%%%%%%%%%%%%%%%%%

% Our standard choice; a very good precision is enforced

options=odeset('RelTol',1e-8,'AbsTol',1e-8);

N =32;
% We consider the case of N = 32

F=2.^[3:11]; 
% Here is a vector of values of the forcing R

meanLyap=zeros(max(size(F)),N); 
% Initialization matrix containing our estimates of the 
% Lyapunov exponents

T=100; 
% Final time; Attention, may need to adapt it to values of R 
% because integrations are more and more expensive as R grows

dt=0.1 
% Step of the Grahm-Schmidt orthogonalization procedure; 
% need to adapt is because for large R lower values are needed 
% or we have no convergence. If so, please define a dt 
% changing with j

n = 1 
% We get an estimate of  Lyapunov exponents every n GS 
% orthogonalizations.

for j=1:max(size(F)); 
    % For loop 

    j   
    % to keep track of things
    
ci=randn(N,1); 
% Random initial condition

tic; %Start clock

[Texpirr,Lexpirr]=lyapunov_mod(N,@lorenz_ext96irr,@ode45,\
0,dt,T,ci,n,options,F(j),N); 
% Computes the Lyapunov exponents for the Lorenz 96 irreversible
% model calling the function lorenz_ext96irr above (to be saved in a 
% separate file as lorenz_ext96irr.m )
% We use a slightly modified version of the freely available 
% lyapunov.m function due to Govorukhin V.N. (2004) see references. 
% We accomodate for the possibily of reading two parameters (F and R  
% in our case). The Benettin et al. (1980) algorithm is used.
%  The routine lyapunov_mod.m is available on request

Lexpirr=Lexpirr(end/2:end,1:end); 
% Estimates of the LEs; let's remove some initial transient behavior

toc 
% Says elapsed time; just to check how long the integration is

meanLyapirr(j,:)=mean(Lexpirr); 
% Mean values of the LEs 

end;  
% end for loop; We have the Lyapunov exponents of the Lorenz 96 
% irreversible model.

%%%%%%%%%%%%%%%% Now to the reversible case

%%%%%%%%%%%%

% Lorenz 96 extended model (trajectory and Jacobian) - 
% Reversible version. Input parameters: N = numbers of
% modes ; R = R parameter

%%%%%%%%%%%%%%%%%%%%%%%%%%%%%%%%%%%%%%%%%%%%%%%%%%%%%%%%%%%%%%%%
%%%save text below as lorenz_ext96rev.m%%%%%%%%%%%%%%%%%%%%%%%%%
function dy=lorenz_ext96rev(t,y,R,N)

%%%%%%%%%  Same as lorenz96rev

dy = zeros(N*(N+1),1);    % a column vector
F=ones(N,1)*R;
Fg2=ones(N,1)*R*sum(y(1:N))/sum(y(1:N).^2)/gamma; 
% Note: this is the fluctuating viscosity 
dy(1)=(y(2)-y(N-1))*y(N)-Fg2(1)*y(1)+F(1);
dy(2)=(y(3)-y(N))*y(1)-Fg2(2)*y(2)+F(2);
dy(N)=(y(1)-y(N-2))*y(N-1)-Fg2(N)*y(N)+F(N);
for j=3:N-1
dy(j)=(y(j+1)-y(j-2))*y(j-1)-Fg2(j)*y(j)+F(j);
end

%%% Initializing the Jacobian 


Y=zeros(N,N);

for j=1:N;
    for k=1:N;
        Y(j,k)=y(N+j+(k-1)*N);
    end;
end;


% Linearized system
% We split the jacobian in two parts

Jac1=zeros(N,N);
Jac2=zeros(N,N);

% Jac1 is itself split into two parts. We are considering the 
% fluctuating viscosity

Jac1a=zeros(N,N);
Jac1b=zeros(N,N);

M=sum(y(1:N));
E=sum(y(1:N).^2)/2;

Jac1a=eye(N)*(-1)*R*M/(2*E);

for j=1:N;
    for k=1:N;
        Jac1b(j,k)=-R*y(j)/(2*E)+2*R*y(j)*y(k)*M/(2*E)^2;
    end;
end;

Jac1=Jac1a+Jac1b;

Jac2(1,2)=y(N);
Jac2(1,N)=y(2);
Jac2(1,N-1)=-y(N);

Jac2(2,3)=y(1);
Jac2(2,N)=-y(1);
Jac2(2,1)=y(3)-y(N);

Jac2(N,1)=y(N-1);
Jac2(N,N-2)=-y(N-1);
Jac2(N,N-1)=y(1)-y(N-2);

for j=3:N-1
Jac2(j,j+1)=y(j-1);
Jac2(j,j-1)=y(j+1)-y(j-2);
Jac2(j,j-2)=-y(j-1);
end

Jac=Jac1+Jac2; %the Jacobian is now obtained
  
%Variational equation   
dy(N+1:N*(N+1))=Jac*Y; 
%Note: MATLAB rearranges correctly the matrix into a vector 
%%%save text above as lorenz_ext96rev.m%%%%%%%%%%%%%%%%%%%%%
%%%%%%%%%%%%%%%%%%%%%%%%%%%%%%%%%%%%%%%%%%%%%%%%%%%%%%%%%%%%
%%%%%%%%%%%%%%%%%%%%%%%%%%%%%%%%%%%%%%%%%%%%%%%%%%%%%%%%%%%%

% Our standard choice; a very good precision is enforced

options=odeset('RelTol',1e-8,'AbsTol',1e-8);

N =32;
% We consider the case of N = 32

F=2.^[3:11]; 
% Here is a vector of values of the forcing R

meanLyaprev=zeros(max(size(F)),N); 
% Initialization matrix containing our estimates of the 
% Lyapunov vectors

T=100; 
% Final time; Attention, may need to adapt it to values of R 
% because integrations are more and more expensive as R grows

dt=0.1 
% Step of the Grahm-Schmidt orthogonalization procedure; 
% need to adapt is because for large R lower values are needed
% or we have no convergence. If so, please define a dt 
% changing with j

n = 1 
% We get an estimate of Lyapunov exponents every n GS 
% orthogonalizations.

for j=1:max(size(F)); 
    % For loop 
    j   
    % to keep track of things
    
ci=randn(N,1); % Random initial condition: Step 1
nor=sum(ci.^2/2/N); ci=ci*sqrt(meanEirr(j))/sqrt(nor); 
% Random initial condition: Step 2: the system is initializated 
% from the energy shell E=meanEirr(j); you must have used Part 1
% of this script

[Texprev,Lexprev]=\ 
lyapunov_mod(N,@lorenz_ext96rev,@ode45,0,dt,T,ci,n,options,F(j),N); 
% Computes the finite-time estimates of the Lyapunov exponents 
% calling the function lorenz_ext96rev above (to be saved in a 
% separate file as lorenz_ext96rev.m )
% We use a slightly modified version of the freely available 
% lyapunov.m function due to Govorukhin V.N. (2004) see references. 
% We  accomodate for the possibily of reading two parameters (F 
% and R in our case).  The Benettin et al (1980) algorithm is used. 
% The routine lyapunov_mod.m is available on request

Lexprev=Lexprev(end/2:end,1:end); 
% Estimates of the LEs; let's remove some initial transient behavior

toc 
% Says elapsed time; just to check how long the integration is

meanLyaprev(j,:)=mean(Lexprev); 
% Mean values of the LEs

end;  
% end for loop; We have the Lyapunov exponents of the Lorenz 96 
% reversible model.
     \end{verbatim}

\vfill\eject
\def\SEC{References}

\end{document}